\documentclass[preprint,review,12pt]{elsarticle}
\usepackage{psfig,epsf,rotating,longtable,txfonts,morefloats}
\usepackage{color}

\journal{Icarus}

\begin{document}
\begin{frontmatter}
\baselineskip=15pt
\textwidth 16.0truecm
\textheight 21.0truecm
\oddsidemargin 0.5truecm
\evensidemargin  0.3truecm
\topmargin 0.2cm
\headsep 1.0cm

\title{Spectral variability on primitive asteroids of the Themis and Beagle families: space weathering effects or parent body heterogeneity?}
\author{S. Fornasier$^{1}$, C. Lantz$^{1}$, D. Perna$^{1}$, H. Campins$^{2}$, M.A. Barucci$^{1}$, D. Nesvorny$^{3}$}

\maketitle
\noindent
$^1$ LESIA, Observatoire de Paris, PSL Research University, CNRS, Univ. Paris Diderot, Sorbonne Paris Cit\'{e}, UPMC Univ. Paris 06, Sorbonne Universit\'es, 5 Place J. Janssen, 92195 Meudon Pricipal Cedex, France \\
$^2$ Department of Physics and Astronomy, University of Central Florida, Orlando, Florida 32816-2385, USA.
$^3$ Department of Space Studies, Southwest Research Institute, Boulder, CO 80302, USA

Submitted to Icarus: July 2015\\
e-mail: sonia.fornasier@obspm.fr; fax: +33145077144; phone: +33145077746\\
Manuscript pages: 40; Figures: 15 ; Tables: 4  \\
\vspace{3cm}

{\bf Running head}: The Themis-Beagle primitive families

{\it Send correspondence to:}\\
Sonia Fornasier  \\
LESIA-Observatoire de Paris  \\
Batiment 17 \\
5, Place Jules Janssen \\
92195 Meudon Cedex \\
France\\
e-mail: sonia.fornasier@obspm.fr\\
fax: +33145077144\\
phone: +33145077746\\ 

\newpage


\begin{abstract}

  Themis is an old and statistically robust asteroid family populating the outer main belt, and resulting from a catastrophic collision that took place 2.5$\pm$1.0 Gyr ago. Within the old Themis family a young sub-family, Beagle, formed less than 10 Myr ago, has been identified. \\
We present the results of a spectroscopic survey in the visible and near infrared range of 22 Themis and 8 Beagle families members. The Themis members investigated exhibit a wide range of spectral behaviors, including asteroids with blue/neutral and moderately red spectra, while the younger Beagle family members look spectrally bluer than the Themis ones and they have a much smaller spectral slope variability. Four Themis members, including (24) Themis, have absorption bands centered at 0.68-0.73 $\mu$m indicating the presence of aqueously altered minerals.\\
  The best meteorite spectral analogues found for both Themis and Beagle families members are carbonaceous chondrites having experienced different degrees of aqueous alteration, prevalently CM2 but also CV3 and CI, and some of them are chondrite samples being unusual or heated. The presence of aqueous altered materials on the asteroids surfaces and the meteorite matches indicate that the parent body of the Themis family experienced mild thermal metamorphism in the past.\\
  We extended the spectral analysis including the data available in the literature on Themis and Beagle families members, and we looked for correlations between spectral behavior and physical parameters using the albedo and size values derived from the WISE data. The analysis of this larger sample confirm the spectral diversity within the Themis family and that Beagle members tend to be bluer and to have an higher albedo. The differences between the two family may be partially explained by space weathering processes, which act on these primitive surfaces in a similar way than on S-type asteroids, i.e. producing reddening and darkening. However we see several Themis members having albedos and spectral slopes similar to the young Beagle members.  Alternative scenarios are proposed including heterogeneity in the parent body having a compositional gradient with depth, and/or the survival of projectile fragments having a different composition than the parent body.
  
\end{abstract}

\end{frontmatter}

\newpage
\section{Introduction}

Themis is one of the most statistically reliable families in
the asteroid belt. First discovered by Hirayama (1918), it has been identified as a
family in all subsequent works, and it has 550 members according to Zappal\'{a} et al. (1995), and more than 5000 members as determined by Nesvorny (2012). 
The Themis family is characterized by asteroids with 3.05 $ \leq a \leq $ 3.22 AU,  0.12 $\leq e \leq$ 0.19, in low inclination orbit $0.7^{o} \leq i \leq 2.22^{o}$ (Zappal\'{a} et al. 1990). It is spectrally dominated by primitive C- and B-type asteroids, as reported by spectroscopic investigation of several members (Moth\'e-Diniz et al. 2005; Florczak et al. 1999; de Leon et al. 2012; Kaluna et al., 2016). The family is old with an estimated age of $\sim$ 2.5$\pm$1.0 Gyr (Bro\v{z} et al. 2013), and it was formed from a large-scale catastrophic disruption of a $\sim$ 270 km (Bro\v{z} et al. 2013) or a $\sim$ 400 km sized parent body (Marzari et al. 1995; Durda et al. 2007). Nesvorny et al. (2008) confirmed that this family is quite old and formed more than 1 Gyr ago.\\
Interestingly, Rivkin \& Emery (2010) and Campins et al. (2010) found spectroscopic evidence of the presence of water ice and organics on (24) Themis. Rivkin and Emery (2010) concluded that its surface contains very fine water frost, probably in the form of surface grain coatings, and that the infrared spectral signatures can be fully explained by a mixture of spectrally neutral material, water ice, and organics. At the same time, Campins et al. (2010) suggested that water ice is evenly distributed over the entire Themis surface using spectra obtained at four different rotational phases. 
Nevertheless the nature of the 3.1 $\mu$m feature on (24) Themis is still a matter of debate, and Beck et al. (2011) proposed the hydrated iron oxide goethite as alternative interpretation of this feature, even if Jewitt \& Guilbert-Lepoutre (2012) stress that goethite, when found in meteorites, is a product of aqueous alteration in the terrestrial environment and that extraterrestrial goethite in freshly fallen meteorites has not been detected. \\
Absorption bands in the visible region related to hydrated silicates have been detected on the surface of several Themis family members (Florczak et al. 1999). These materials are produced by the aqueous alteration process, that is a low temperature ($<$ 320 K) chemical alteration of materials by liquid water (Vilas \& Sykes, 1996). The presence of hydrated minerals implies that liquid water was present on these asteroids in the past, and suggests that post-formation heating took place.
The  main belt comets 133P/Elst-Pizarro, 176P/LINEAR, 288P/(300163) 2006 VW139, and possibly 238P/Read may be related to the Themis family (Nesvorny et al. 2008; Hsieh et al. 2006, 2009, 2012; Haghighipour 2009) because of orbital proximity and spectral properties analogies. According to Licandro et al. (2011, 2013) works, the visible spectra of the main belt comets 133P, 176P, and 288P  are compatible with those of Themis and Beagle families members.
In particular Nesvorny et al. (2008) propose that 133P could potentially be one member of the Beagle sub-family within the Themis group. This sub-family is a young cluster with an estimates age $< 10$ Myr (Nesvorny et al. 2008).

The detection of water ice on asteroid (24) Themis, of hydrated silicates on several Themis members and the spectral and dynamical link between the Themis family and some main belt comets all indicate that Themis family is an important  reservoir of water in the outer main belt. These discoveries support the suggestion that at least some of Earth's current supply of water was delivered by asteroids some time following the collision that produced the Moon (Morbidelli et al. 2000, Lunine 2006). \\
With this paper we investigate both regular Themis family members (22) and younger ones (8), belonging to the Beagle sub-family, aiming to constrain their surface composition, to establish spectral links with meteorites, and to investigate potential space weathering effects and spectral variability between the two families members.

\section{Observations and data reduction} 

Spectroscopic observations in the visible and near infrared range of Themis and Beagle families members were made at the 3.56 m Italian Telescopio Nazionale Galileo (TNG) of the European Northern Observatory (ENO) in la Palma, Spain, in two runs, on 16-20 February and 16-18 December 2012.
For visible spectroscopy we used the Dolores (Device Optimized for the 
LOw RESolution) instrument equipped with the low resolution red (LRR) and blue (LRB) grisms.
The LRR grism covers the 0.52--0.95 $\mu$m range with a spectral dispersion 
of 2.9 \AA/px, while the LRB grism covers the 0.38-0.80 $\mu$m range with a spectral dispersion 
of 2.8 \AA/px. 
The Dolores detector is a 2048 $\times$ 2048 pixels Loral thinned and 
back-illuminated CCD, having a pixel size of 
15 $\mu$m and a pixel scale of 0.275 arcsec/px. 
The red and blue spectra in the visible range were separately reduced and finally combined together to obtain spectral coverage from 0.38 to 0.95 $\mu$m. Like most of the Loral CCDs, the Dolores chip is affected by moderate-to-strong fringing
at red wavelengths. Despite taking as much care as possible in
the data reduction process, some asteroid spectra acquired with the LRR grism show residual
fringing that impedes identification of absorption bands and an accurate analysis in the 0.81-0.95 $\mu$m range when NICS observations are not available. \\
For the infrared spectroscopic investigation  we used the
near infrared camera and spectrometer (NICS) equipped with an Amici prism disperser. 
This equipment covers the 0.78--2.40 $\mu$m range during a single 
exposure with a spectral resolution of about 35 (Baffa et al. 2001). 
The detector is a 1024 $\times$ 1024 pixel Rockwell HgCdTe Hawaii array. 
The spectra were acquired nodding the object
along the spatial direction of the slit, in order to obtain alternating pairs 
(named A and B) of near--simultaneous images for the  background removal. 
For both the visible and near infrared observations we utilized a 2 arcsec 
wide slit, oriented along the parallactic angle to minimize 
the effect of atmospheric differential refraction.

Visible and near-infrared spectra were reduced using ordinary procedures 
of data reduction with the software packages Midas as described in
Fornasier et al. (2004, 2008). 
For the visible spectra, the procedure includes the subtraction of the bias 
from the raw data, flat--field correction, cosmic ray removal, sky subtraction,
collapsing the two--dimensional spectra to one dimension, wavelength calibration,
and atmospheric extinction correction, using La Palma atmospheric extinction coefficients.
The spectra were normalized at 5500 \AA. 
The reflectivity of each asteroid was 
obtained by dividing its spectrum by that of the solar analog star closest in
time and airmass to
the object. Spectra were finally smoothed with a median filter 
technique, using a box of 19 pixels in the spectral direction for each point of
the 
spectrum. The threshold was set to 0.1, meaning that the original value was
replaced by 
the median value if the median value differs by more than 10\% from the original
one.

\begin{figure*}
\centerline{\psfig{file=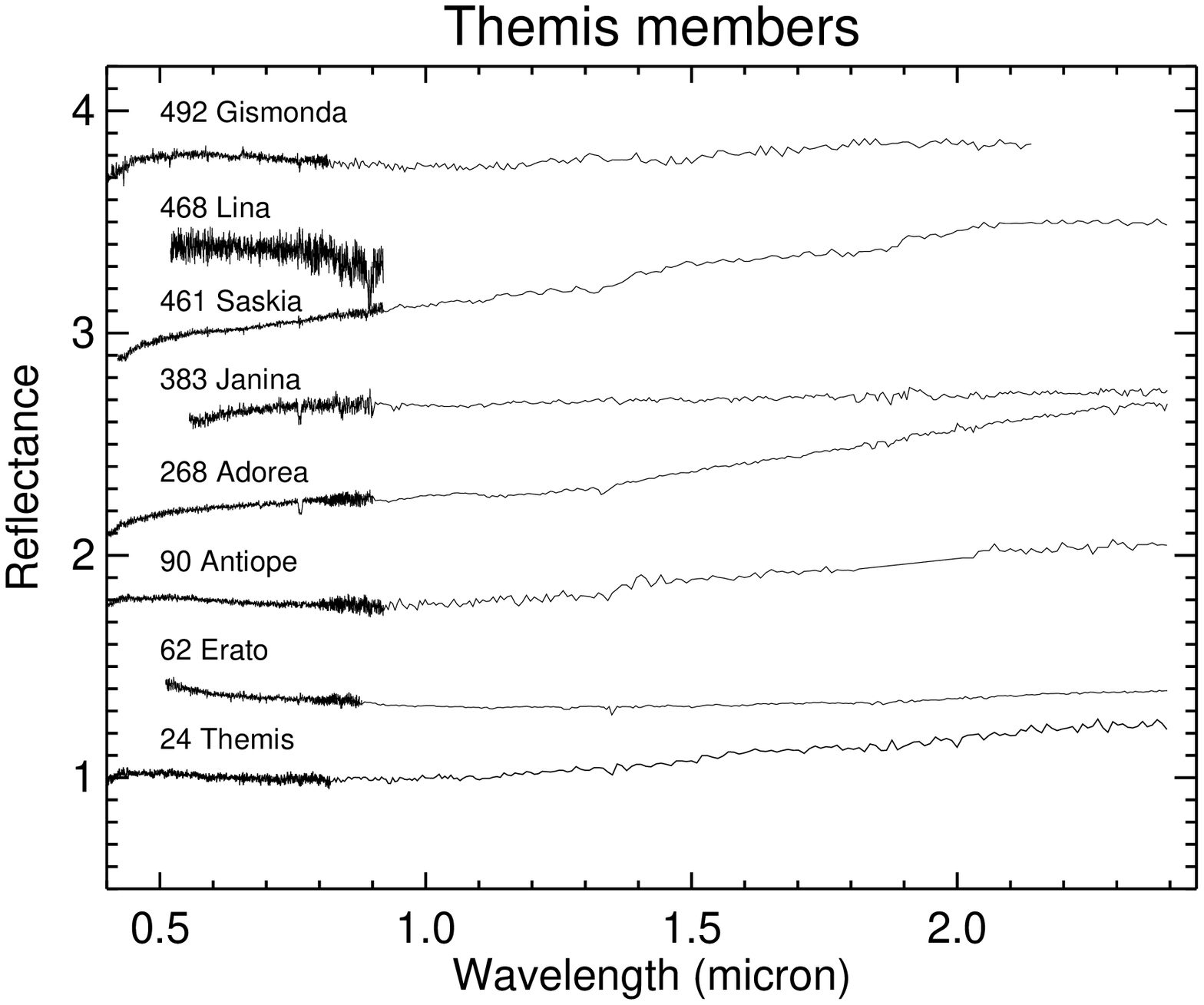,width=20truecm,angle=0}}
\caption{Reflectance spectra of the Themis family members investigated. Spectra are shifted by 0.4 in reflectance for clarity.}
\label{f1}
\end{figure*}

\begin{figure*}
\centerline{\psfig{file=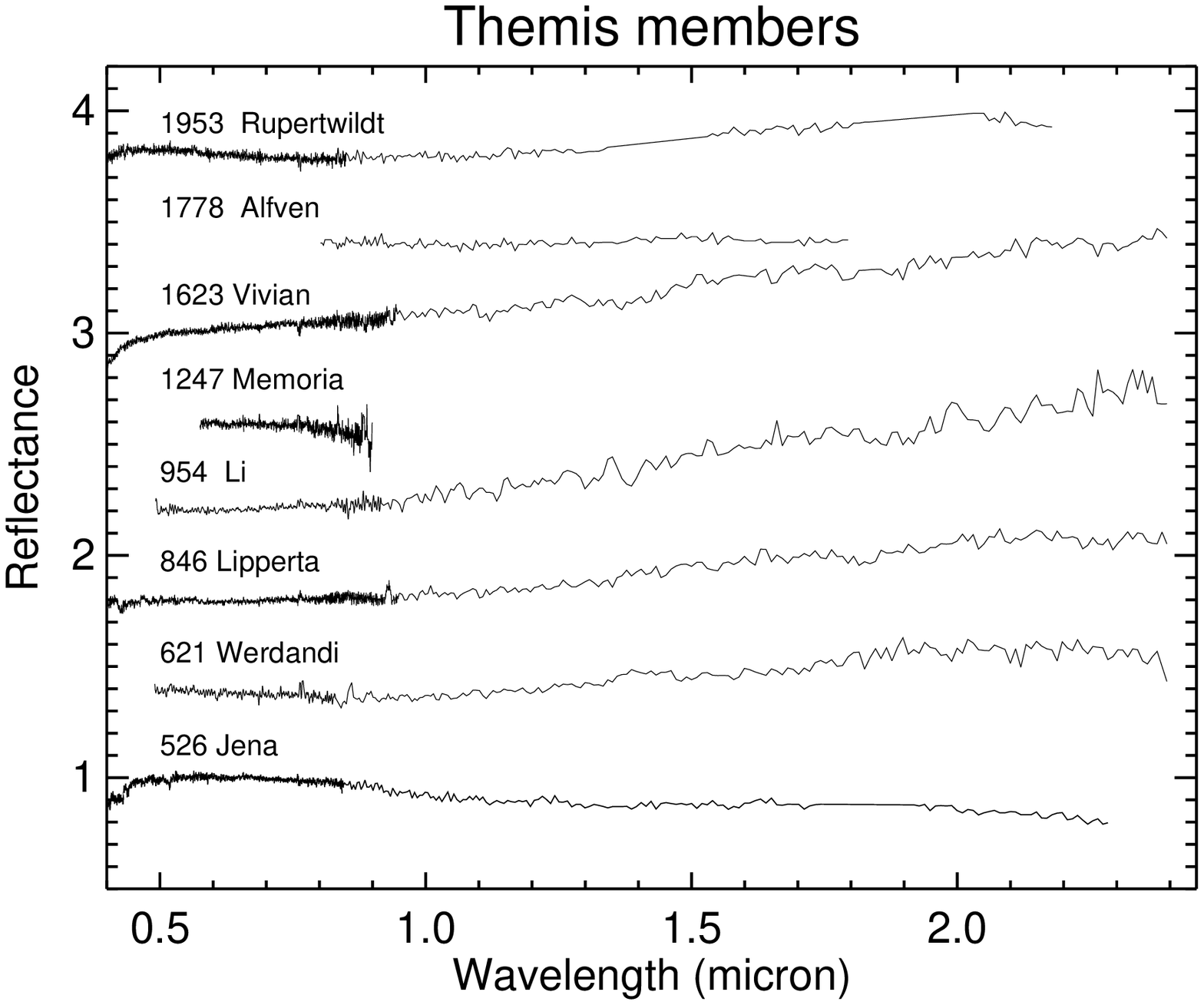,width=20truecm,angle=0}}
\caption{Reflectance spectra of the Themis family members investigated. Spectra are shifted by 0.4 in reflectance for clarity.}
\label{f2}
\end{figure*}

\begin{figure*}
\centerline{\psfig{file=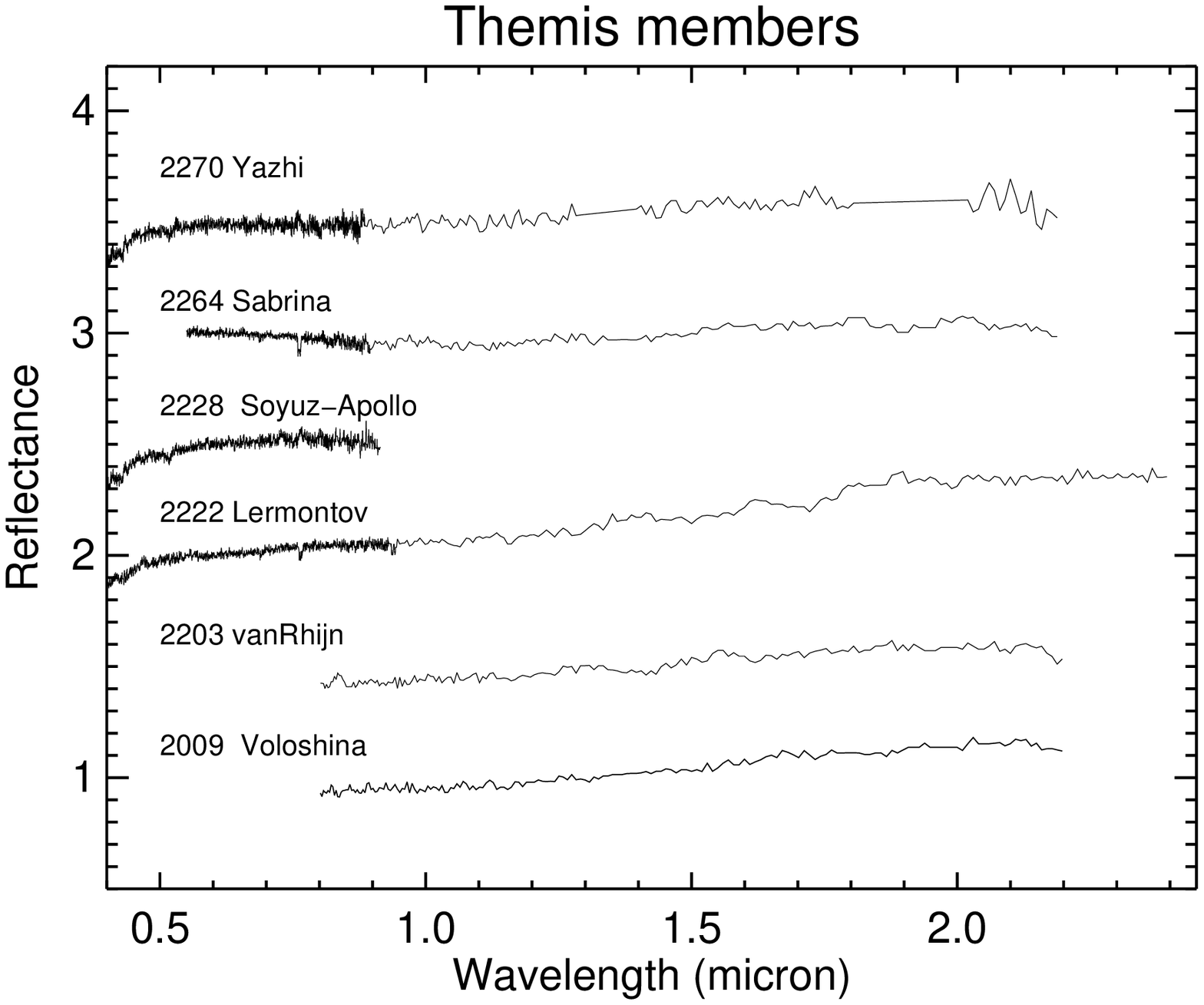,width=20truecm,angle=0}}
\caption{Reflectance spectra of the Themis family members investigated. Spectra are shifted by 0.5 in reflectance for clarity.}
\label{f3}
\end{figure*}

\begin{figure*}
\centerline{\psfig{file=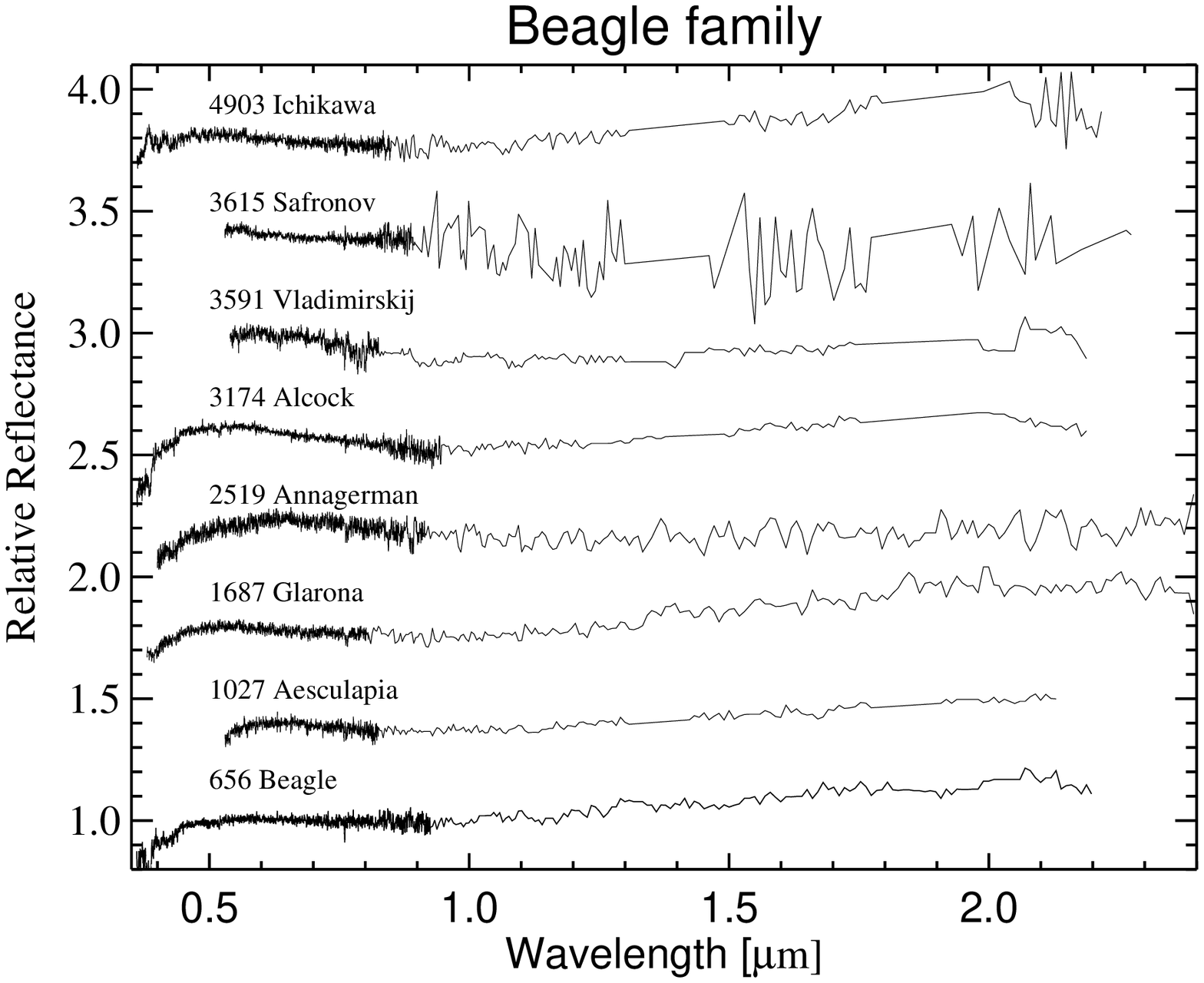,width=20truecm,angle=0}}
\caption{Reflectance spectra of the Beagle family members investigated. Spectra are shifted by 0.4 in reflectance for clarity.}
\label{f4}
\end{figure*}

For observations in the infrared range, spectra were first corrected for flat
fielding, 
then bias and sky subtraction
was performed by producing A-B and B-A frames. The positive spectrum of the B-A
frame was 
shifted and added to the positive spectrum of 
the A-B frame. The final spectrum is the result of the mean of all pairs of 
frames previously combined. The spectrum was extracted and wavelength 
calibrated. Due to the very low resolution of the Amici prism, 
the lines of Ar/Xe lamps are blended and cannot be easily used for standard reduction procedures. 
For this reason, the wavelength calibration was obtained using a look-up table
which is  based on the theoretical dispersion predicted by ray-tracing and adjusted 
to best fit the observed spectra of calibration sources.  Finally, the extinction correction and solar removal was obtained by 
division of each asteroid spectrum by that of the solar analog star closest in
time and airmass to the object. The stellar and asteroid spectra
were cross-correlated and, if necessary, sub-pixel shifts were
made before the asteroid-ratio star division was done. \\
The infrared and visible spectral ranges of each asteroid were finally
combined by overlapping the spectra, merging the two wavelength
regions at the common wavelengths and utilizing the zone
of good atmospheric transmission to find the normalization factor
between the two spectral parts. The
overlapping region goes from about 0.78-0.92 $\mu$m. For most of the data  we took the
average value in reflectance of the visible spectrum in the 0.89--0.91 $\mu$m region 
to normalize the reflectance of the infrared spectrum. For the spectra where LRR grism data were not available or present but with strong fringing problems, we joint the visible and NIR data using the mean reflectance value of the visible spectrum in the 0.79-0.81 $\mu$m region. \\
Asteroids having both visible and NIR data or only visible spectra were {\bf finally all normalized in relative reflectance at 1 at the wavelength 0.55 $\mu$m}, while those observed only in the NIR region were normalized  at 1.25 $\mu$m. 
Details on the observing conditions are reported in Table~\ref{tab1}, and the spectra of the observed asteroids are shown in Figs.~\ref{f1}-\ref{f4}.

[Here Figs 1, 2, 3, 4]

[Here Table 1]

%
                          \section{Results}
%

                          [Here Table 2 and Figure 5]

We present new spectra of 8 Beagle and 22 Themis families members.
Most of the asteroids were observed in the V+NIR range, except for some targets for which
we did not get the visible spectra (objects 621, 954, 1778, 2009, and 2203; for asteroids 621 and 954 we complete our data with the visible spectra from the
$S^3S0^2$ survey (Lazzaro et al. 2004)), or the NIR ones (62, 383, 268, 526, 1247, and 2228; for objects 62 and 383 we got the IR spectra from Clark et al. (2010)). 
The observed asteroids and their proper elements are reported in Table~\ref{slope}. \\
The members of each family were derived from the family lists from analytic proper elements
(Nesvorny et al. 2012).  A hierarchical clustering method has been applied to the asteroids proper elements to identify the dynamical families (Nesvorny et al., 2008). Themis is an old and huge family with 5425 members identified at a cutoff velocity of 60 m/s (Nesvorny et al. 2012). Beagle is a young smaller cluster within Themis, having 149 members at a cutoff velocity of 20 m/s, and an estimated age $<$ 10Myr (Nesvorny et al. 2008, 2012).  The cutoff velocity is related to the relative velocity of the fragments as they were ejected from the sphere of influence of the parent body, and lower values implies a higher statistical significance of the family. This means that the Beagle family is very robust.
        
\begin{figure*}
\centerline{\psfig{file=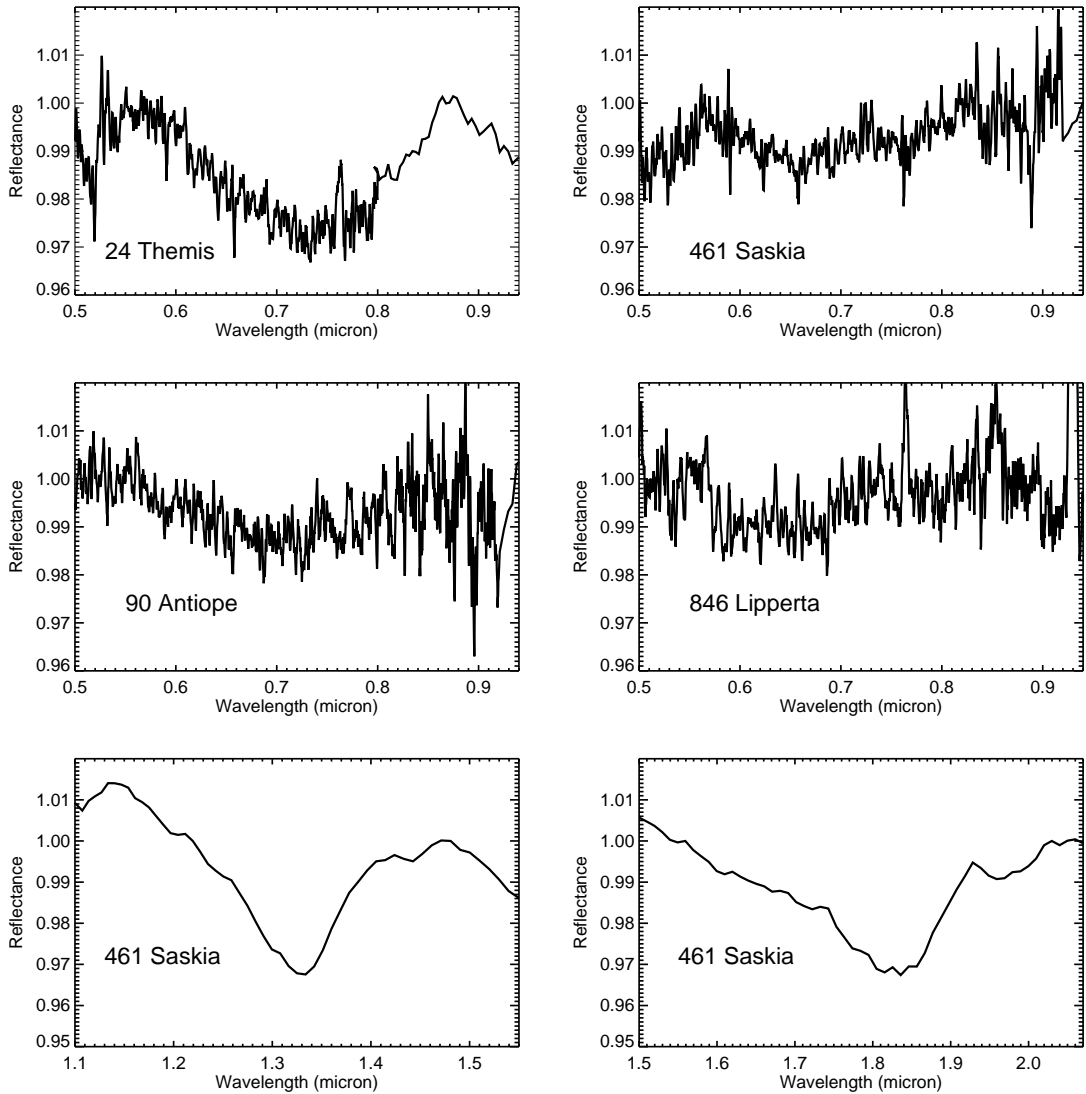,width=15truecm,angle=0}}
\caption{Continuum removed spectra of the 4 asteroids showing absorption features. The visible spectrum of (24) Themis is the one acquired on 19 December.}
\label{f5bis}
\end{figure*}

All the objects investigated belong to the C or B spectral types, following the Tholen (1984) classification scheme. The majority of the spectra are featureless, and in particular none of the investigated spectra show water ice absorption bands at 1.5 and 2 micron. A few Themis members ((24) Themis, (90) Antiope, (461) Saskia, and (846) Lipperta) show weak spectral absorption features in the visible range associated with hydrated minerals (Table~\ref{hydra}, and Fig.~\ref{f5bis}), indicating that they have experienced the aqueous alteration process, as already found for other Themis family members (Florczak et al. 1999).
We considered only the absorption features deeper than the peak-to-peak  scatter (that is depth $>$ 0.8\%) in the spectrum, and we characterized them following the method described in Fornasier et al. (1999, 2014). This method includes first a smoothing of the spectra using a box of 6 pixels in the spectral direction, then the evaluation of the linear continuum at the edges of the band identified, the division of the original spectrum by the continuum, and the fitting  of the absorption band with a polynomial of order 2--4.  The band center was then calculated as the position where the minimum of the spectral reflectance curve occurs (on the polynomial fit), and the band depth as the minimum of the polynomial fit. The main absorption band identified is the one centered in the 0.68-0.73 $\mu$m region which is attributed to $Fe^{2+}\rightarrow Fe^{3+}$ charge transfer absorptions in phyllosilicate minerals (Vilas \& Gaffey 1989; Vilas et al. 1993). This band is often associated with an evident UV absorption below $\sim$ 0.5 $\mu$m, due to a strong ferric oxide intervalence charge transfer transition (IVCT) in oxidized iron, and is often coupled with
other visible absorption features related to the presence of aqueous alteration products (e.g. phyllosilicates, oxides, etc) (Vilas et al. 1994). We see the UV drop-off in reflectivity in most of the asteroids observed with the LRB grism, that is on Themis members (24) Themis, (268) Adorea, (461) Saskia, (492) Gismonda, (526) Jena, (1623) Vivian, (222) Lermontov, (2228) Soyuz-Apollo, (2270) Yazhi, and on Beagle members (656) Beagle, (1687) Glarona, (2519) Annagerman, (3174) Alcock, and (4903) Ichikawa,  but not on (90) Antiope and (846) Lipperta, that have a faint absorption centered at $\sim$ 0.7 $\mu$m. These IVCTs comprise multiple absorptions that are not uniquely indicative of phyllosilicates, but are present in the spectrum of any object
containing Fe$^{2+}$ and Fe$^{3+}$ in its surface material (Vilas 1994). Thus, even if most of these spectra are featureless in the VIS-NIR range, their UV drop-off in reflectivity may indicate the presence of phillosilicates (Vilas 1994).  \\ 
Saskia also shows two absorption features in the NIR region centered at about 1.32 $\mu$m and 1.82 $\mu$m, as determined by a 4$^{th}$ order polynomial fit to the absorption bands. However, the center of these bands fall in the region of very low atmosphere transmission so their center position cannot be precise and their identification needs to be confirmed by other observations. If real, these absorptions look similar to those found on some hydrous Fe-bearing hydrated silicates (Bishop \& Pieters 1995), thus potentially associated to the presence of hydrated minerals.  The continuum removed spectra of asteroids showing faint absorption features are shown in Fig.~\ref{f5bis}. \\
We identified features associated to hydrated silicates only on Themis members. Thus only 18\% of the 22 Themis members investigated show evidence of aqueous alteration in the visible and near infrared ranges. Recently Kaluna et al. (2016) investigated a sample of 22 Themis and 23 Beagle members. The asteroids they observed are all different from those of our survey, and they are all smaller than 15 km. Interestingly their analysis also indicates a lack of the 0.7 $\mu$m features in the observed asteroids, with only one Beagle member having that absorption. It must be noted that the absence of the 0.7 $\mu$m feature does not necessarily imply a lack of aqueous alteration, considering that this absorption is much fainter than the main absorption feature associated to hydrated minerals and centered at 3 $\mu$m (Lebofsky 1980; Jones et al. 1990), and that not all the aqueous altered  CM/CI meteorites show this feature (McAdam et al. 2015). Considering the results of other surveys,  Florczak et al. (1999) found  absorption bands associated to hydrated silicates in 15 out of 36 Themis members observed in the visible range, while Fornasier et al. (2014, Table 4), observed the $\sim$ 0.7 $\mu$m band only on 8 out of 47 Themis members spectra.\\
Two out of the three Themis members observed by Licandro et al. (2011) are in common with our observations: (62) Erato and (383) Janina. For (62) Erato Licandro et al. found a slope value (-2$\pm$1 \%/(1000\AA) similar to the one we found (Table~\ref{slope}), while the Janina spectrum is quite different, having a positive visible slope in our data and a negative one in both Licandro et al. (2011) and  Bus \& Binzel (2003) observations. Additional data covering different rotational phases are needed to cast light on potential surface heterogeneities of this asteroid.    

[Here Table 3]

\subsection{(24) Themis: surface heterogeneities}

[Here FIGURE 6]

\begin{figure*}
\centerline{\psfig{file=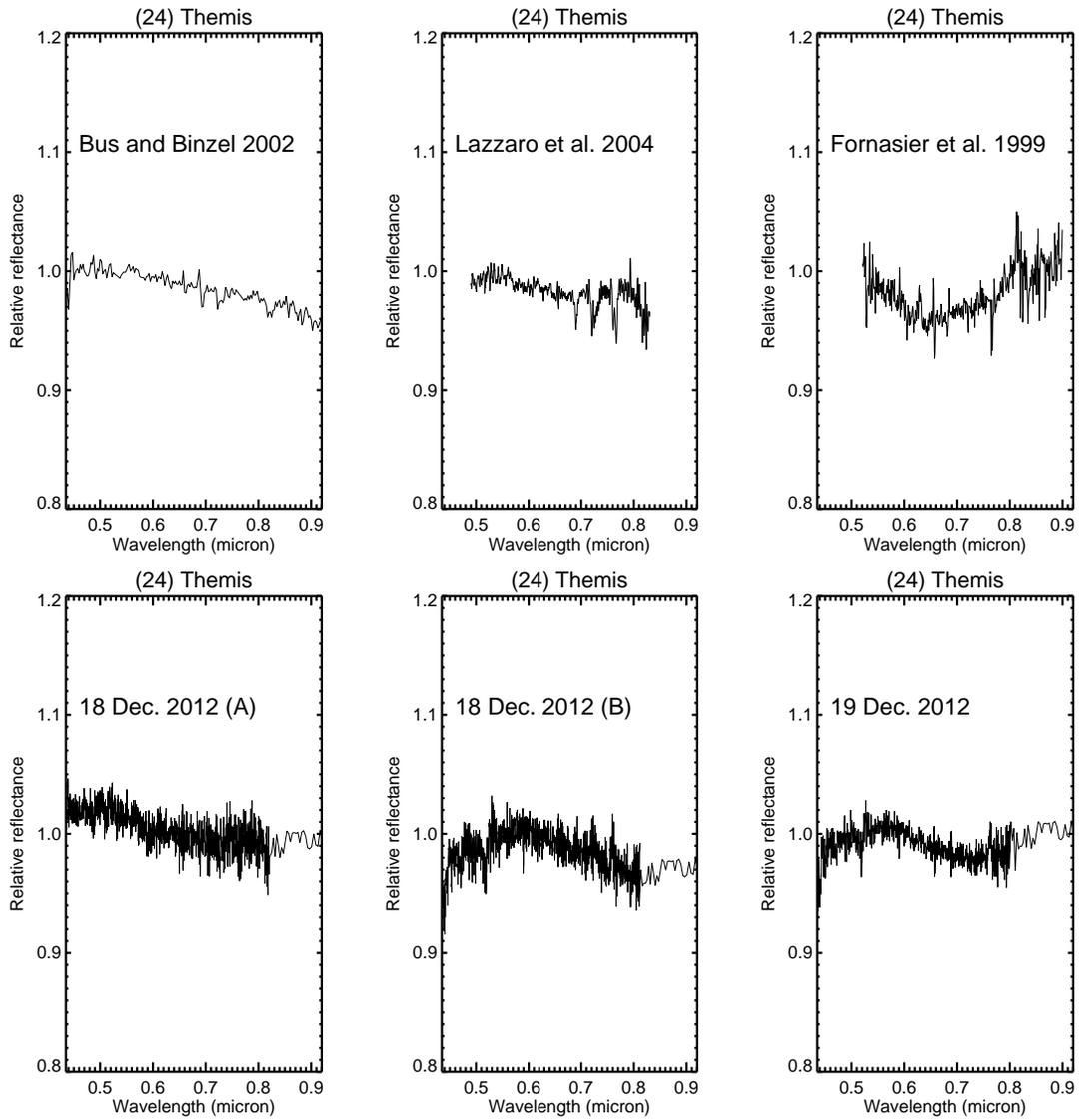,width=15truecm,angle=0}}
\caption{Relative reflectance spectra in the visible region of (24) Themis taken at different aspect and rotational phases from data presented here and from the literature. The asteroid shows surface heterogeneities. }
\label{f5}
\end{figure*}

(24) Themis is a big asteroid with a diameter of  218$\pm$1 km and a low albedo value of  6$\pm$1\% (Hargrove et al. 2015).  
We observed (24) Themis three times in the visible range and once in the NIR range during the December 2012 run, finding some spectral variability (Fig.~\ref{f5}). Unfortunately  all the LRR spectra of (24) Themis  suffered of strong fringing problems, and they have a very bad S/N ratio in the 0.8-0.95 $\mu$m. We thus complete the {\bf 18 and 19 Decem} LRB observations with the available NICS spectrum, that has a better S/N ratio in 0.8-0.95 $\mu$m range than the LRR data. The spectra acquired on 18 Dec. have a negative slope in the visible range and no hints of absorption bands, while the one taken on 19 Dec. shows an absorption band centered at $\sim$ 7333$\pm$48\AA, having a depth of 2.6\%, attributed to hydrated silicates. 
The data also indicate a clear UV-drop-off of the reflectance for wavelength $<$  0.5 $\mu$m for the observations acquired on 18 Dec. UT 22 and 19 Dec. Assuming a rotational period of 8.37677$\pm$0.00002 h (Higley, 2008) and taking the LRB spectrum of 18 Dec. UT 02:16 as reference for the rotational phase, then the second and third LRB spectra of (24) Themis fall at rotational phases of 0.3020 and 0.7278, respectively. We thus covered different rotational phases and the observed spectral differences may be related to surface heterogeneities on (24) Themis. This conclusion is supported also by the comparison with (24) Themis spectra available in the literature (Fig.~\ref{f5}), taken from Fornasier et al. (1999), Bus\& Binzel (2002), and Lazzaro et al. (2004). In particular Fornasier et al. (1999) reported an absorption feature attributed to hydrated silicates, similar to the one found in this paper but centered at a shorter wavelength (6722$\pm$39 \AA), with a depth of 3.5\% compared to the continuum, while the other literature spectra are featureless except for the potential presence of the UV-drop-ff in reflectivity.

(24) Themis shows strong evidence of surface water ice (Campins et al. 2010; Rivkin et al. 2010), that seems to be widespread present on its surface, according to Campins et al. (2010) results. Its emissivity spectrum  exhibits a rounded 10 $\mu$m emission feature, found also on other Themis family members (Licandro et al. 2012), attributed to small silicate grains embedded in a relatively transparent matrix, or from a very under-dense surface structure (Hargrove et al. 2015). Recently McAdam et al. (2015) interpret the emissivity behavior of (24) Themis as due to complex surface mineralogy with approximately 70 vol.\% phyllosilicates and 25 vol.\% anhydrous silicate. \\
Considering our data and those of the literature, (24) Themis most likely has an heterogeneous surface composition, presenting both water ice and hydrated silicates together with dark phases like those found on carbonaceous chondrites. According to Castillo-Rogez \& Schmidt (2010) models, large Themis family objects may contain a large fraction of the parent body ice shell, and this ice may not be completely pure, but possibly contaminated with other materials such as organics, oxides and  hydrated minerals. Jewitt \& Guilbert-Lepoutre (2012) report no detection of gas from sublimated ice on (24) Themis, and conclude that ice must cover a relatively small fraction (10\%) of the asteroid surface, having potentially been exposed on the surfaces of (24) Themis by recent impacts. Surface heterogeneity seems to be common on the largest asteroids/dwarf planet imaged so far, as seen for Vesta and Ceres by the Dawn mission. For instance several bright spots are observed on Ceres, and potentially attributed to water ice associated with impact craters (Reddy et al. 2015), and groundbased observations underline spectral variability on its surface (Perna et al. 2015).\\
Future observations  with high S/N rotationally resolved spectra in the visible and near infrared region are needed to fully investigate the (24) Themis surface heterogeneity.

\subsection{Spectral slopes and physical parameters}

[Here Figs. 7 and 8]

To analyze the data, we compute spectral slope values with linear
fits to different wavelength regions: $S_{cont}$ is the spectral slope in the whole
range observed for each asteroid, $S_{VIS}$ is the slope in the 0.55-0.80 $\mu$m
range, S$_{NIR1}$ is the slope in the 0.9-1.4 $\mu$m range, S$_{NIR2}$ is
the slope in the 1.4-2.2 $\mu$m range, and  S$_{NIR3}$ is
the slope in the 1.1-1.8 $\mu$m range. Values are reported in
Table~\ref{slope}. The slope error bars take into account the
$1\sigma$ uncertainty of the linear fit plus 0.5\%/$10^3$\AA\ attributable to the spectral variation due to the use of different solar analog stars during the night. \\

[Here Figs. 9 and 10]

\begin{figure*}
\centerline{\psfig{file=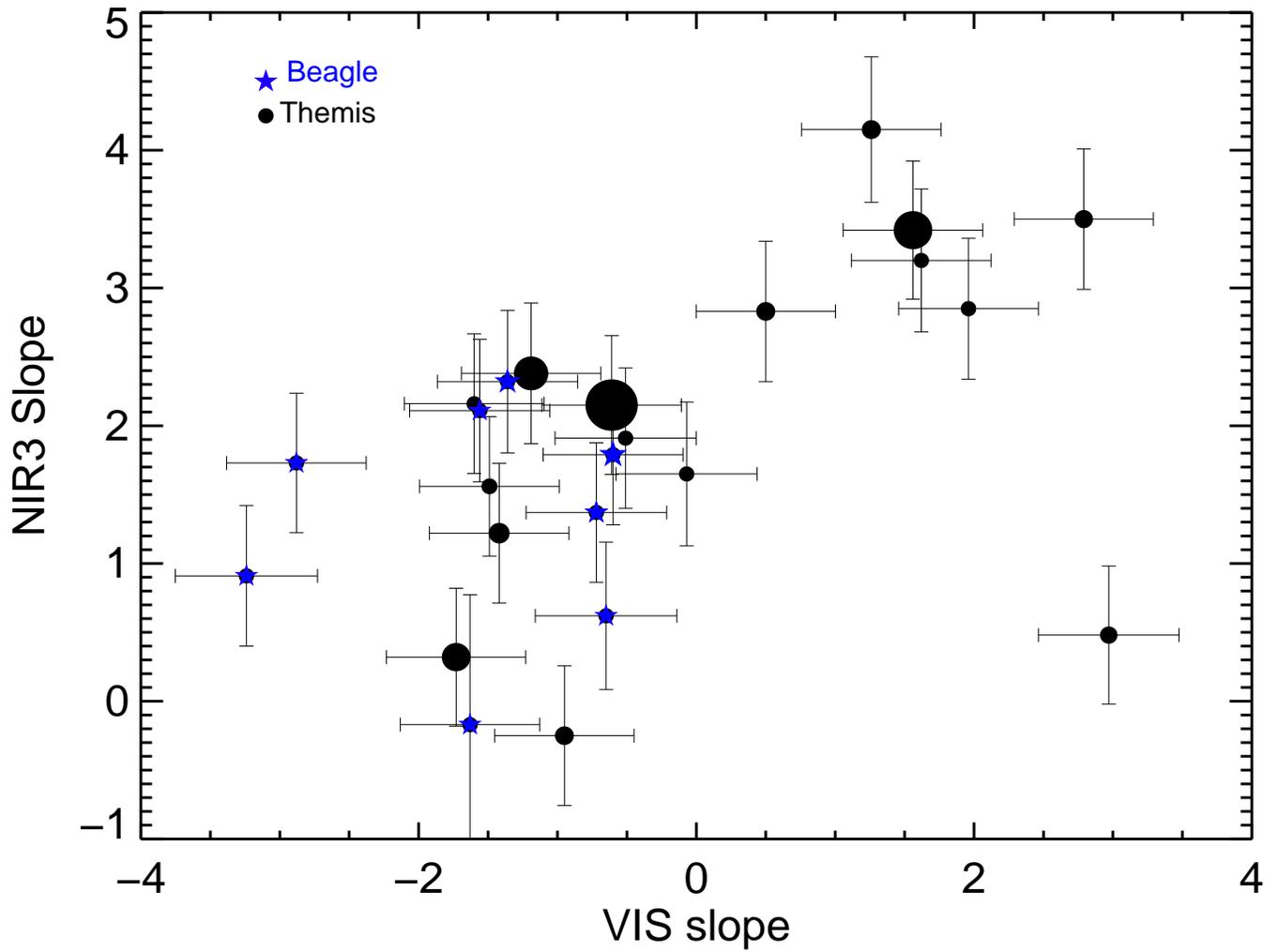,width=20truecm,angle=0}}
\caption{NIR3 spectral slope (1.1-1.8 $\mu$m) versus VIS spectral slope (0.5-0.8 $\mu$m) for the Themis (black circles) and Beagle (blue stars) members. The spectral slopes are in \%/(10$^{3}$\AA). The size of the circles is proportional to the asteroids' diameters.}
\label{f6}
\end{figure*}

Our observations clearly show a range of spectral behaviors exhibited by Themis family members in the visible and near-infrared (Figs.~\ref{f1}-~\ref{f3}, and~\ref{f6}), including asteroids with blue/neutral and moderately red spectra (relative to the Sun). Our findings are consistent with, and complement previous spectroscopic studies that hinted at spectral diversity within the Themis family in the visible (Florczak et al. 1999; Kaluna et al., 2016), the near-infrared (Ziffer et al. 2011; de Le\'on et al. 2012; Al\'i-Lagoa et al. 2013), and the mid-infrared (Licandro et al. 2012). On the other hand, the robust and younger Beagle family members look different, with a smaller spectral slope variability. This family seems dominated by objects with a blue/neutral spectrum in the visible range, and with a neutral to moderately red spectral behavior in the near infrared range.

\begin{figure}
\centerline{\psfig{file=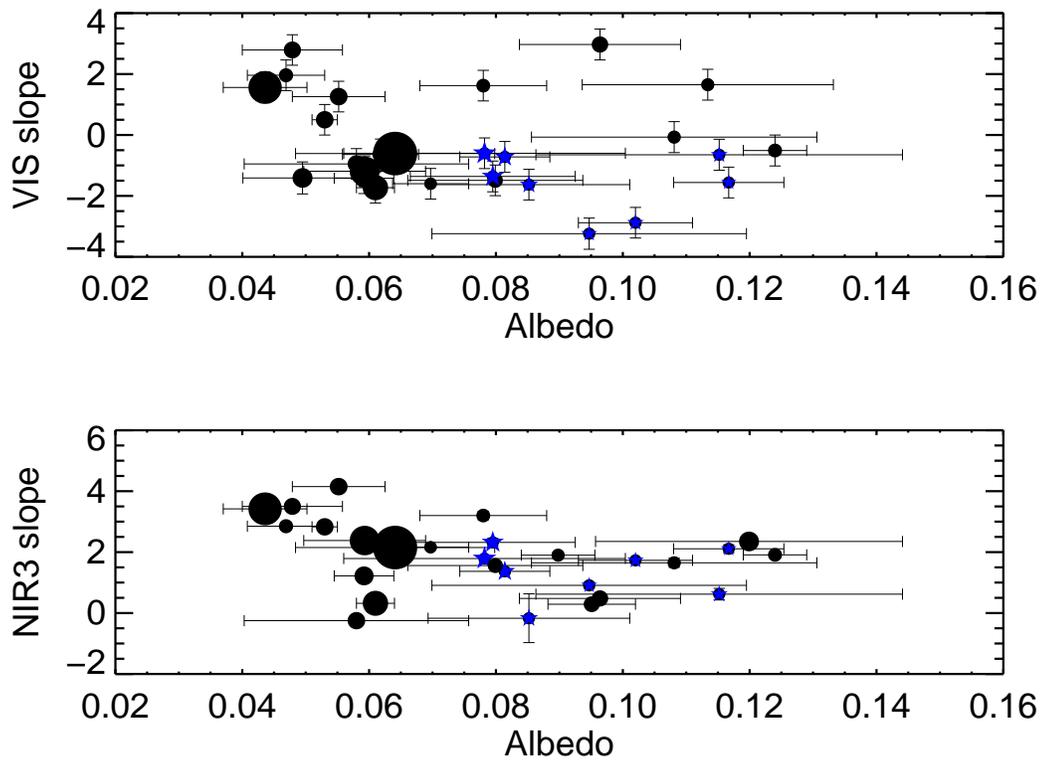,width=15truecm,angle=0}}
\caption{Spectral slopes in the visible (0.55-0.8 $\mu$m) and in the NIR (1.1-1.8$\mu$m) ranges versus albedo. The spectral slopes are in \%/(10$^{3}$\AA). Beagle and Themis family members are represented with blue stars and black circles, respectively. The size of the circles is proportional to the asteroids' diameters.}
\label{f7}
\end{figure}

\begin{figure}
\centerline{\psfig{file=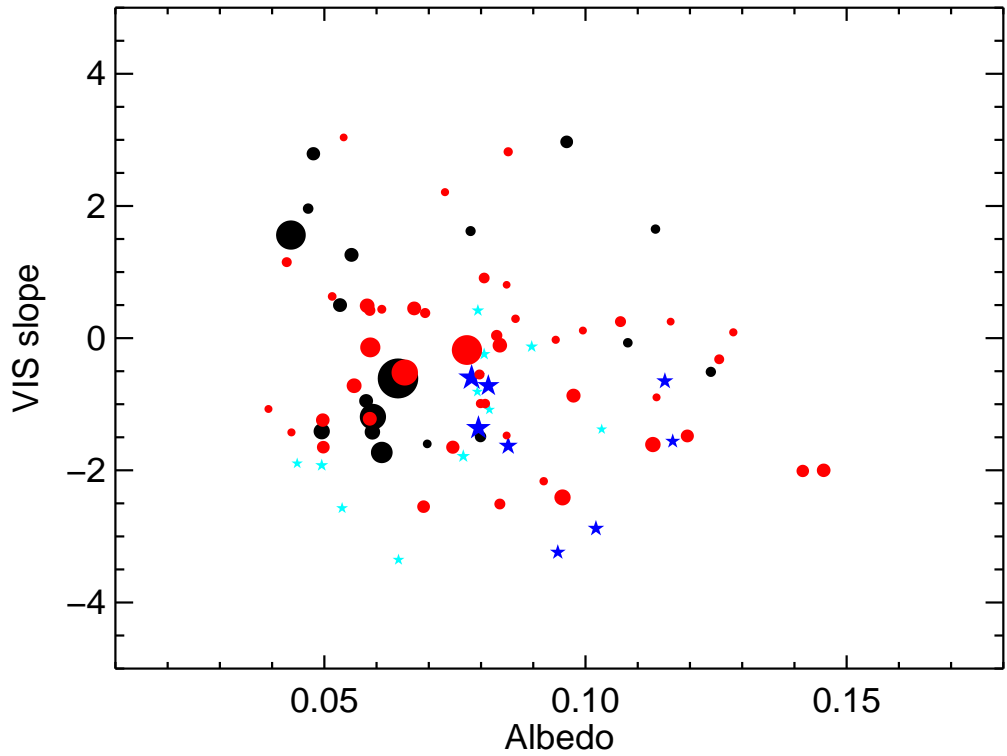,width=15truecm,angle=0}}
\caption{Spectral slope in the visible wavelength range (0.55-0.8 $\mu$m, units \%/(10$^{3}$\AA)) versus the albedo for the Themis and Beagle members. Beagle family members are represented with stars, blue from our data and cyan from the literature data, while the Themis members are represented with circles, black from our data and red from the literature data. The size of the symbols is proportional to the asteroids' diameters.}
\label{f8}
\end{figure}

\begin{figure}
\centerline{\psfig{file=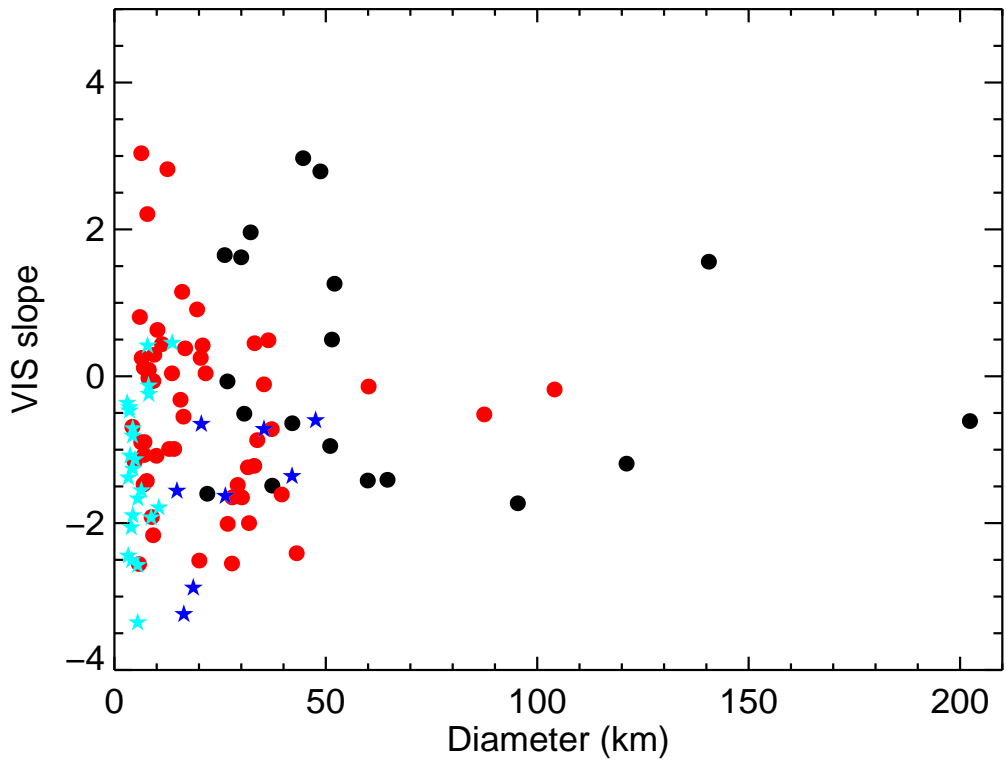,width=15truecm,angle=0}}
\caption{Spectral slope in the visible wavelength range (0.55-0.8$\mu$m, units \%/(10$^{3}$\AA)) versus asteroids' diameter. Beagle family members are represented with star symbol, blue from our data and cyan from the literature data,  while the Themis members are represented as circles, black from our data and red from the literature data.}
\label{f9}
\end{figure}

To better understand the difference between the two families, we look for correlations between spectral slopes and physical parameters such as the albedo and diameter, which were derived from the WISE data (Masiero et al. 2011).
Figure~\ref{f7} shows the visible and NIR3 spectral slopes versus the WISE albedo ($p_v$) for the Themis and Beagle families members here investigated. It clearly appears that the young Beagle members are not only spectrally bluer but tend to have a higher albedo value (mean $p_v$ = 0.0941$\pm$0.0055) than most of the Themis members investigated (mean $p_v$ = 0.0743$\pm$0.0054). \\
To test if this trend is true, we extended the analysis on a larger sample including 79 Themis and Beagle families members spectra available in the literature from several surveys in the visible wavelength range (Xu et al. 1995; Bus \& Binzel 2002; Lazzaro et al. 2004; Fornasier et al. 1999, 2014; Kaluna et al., 2016). For most of these data, the spectral slope has been evaluated in a similar way to what is done here, i. e. between 0.55-0.8 $\mu$m. Only the spectral slope for asteroids observed in Kaluna et al. (2016) was evaluated in a different manner, between 0.49 and 0.91 $\mu$m. However, beeing their spectra normalized at the same wavelength than us and featureless, the variation in the spectral slope values associated to the  different wavelength ranges used must be minimal. For 11 out of 23 Beagle members and 47 out of 56 Themis members of the literature data the albedo value is available from the WISE observations. Figure~\ref{f8} shows the visible spectral slope versus albedo of this extended sample. The trend seen in our data is confirmed and it is evident that there are no Beagle members with a red spectral slope.  Conversely, Themis members show an important spectral variety, having both blue and moderately red slopes and low to moderate albedo values (4-15\%). This spectral variety is not related to different sizes of the asteroids investigated because Themis members having diameter similar to the Beagle ones span different VIS spectral slopes (-2.5 $\%/(10^{3}$\AA) to 3 $\%/(10^{3}$\AA)) and albedo values (4-12.5\%), as shown in Figure~\ref{f9}. \\
The difference in the albedo values is less pronunced, Themis members having a slightly lower mean albedo  (0.0784$\pm$0.0031) than Beagle ones  (0.0819$\pm$0.0045). However, analyzing the average albedo from WISE data on a larger sample and comparing members of similar sizes,  Kaluna et al. (2016) found that the small Themis members ($D <$ 15km) have a lower albedo ($p_v$=0.068$\pm$0.001) than the larger asteroids ($p_v$=0.075$\pm$0.001) of the family, and that their albedo is  significantly lower than that of the Beagle population ($p_v$=0.079$\pm$0.005). Their results clearly indicate that Beagle members have a sligthly higher albedo value than the Themis members of similar size.

[Here Fig. 11]

\begin{figure*}
\centerline{\psfig{file=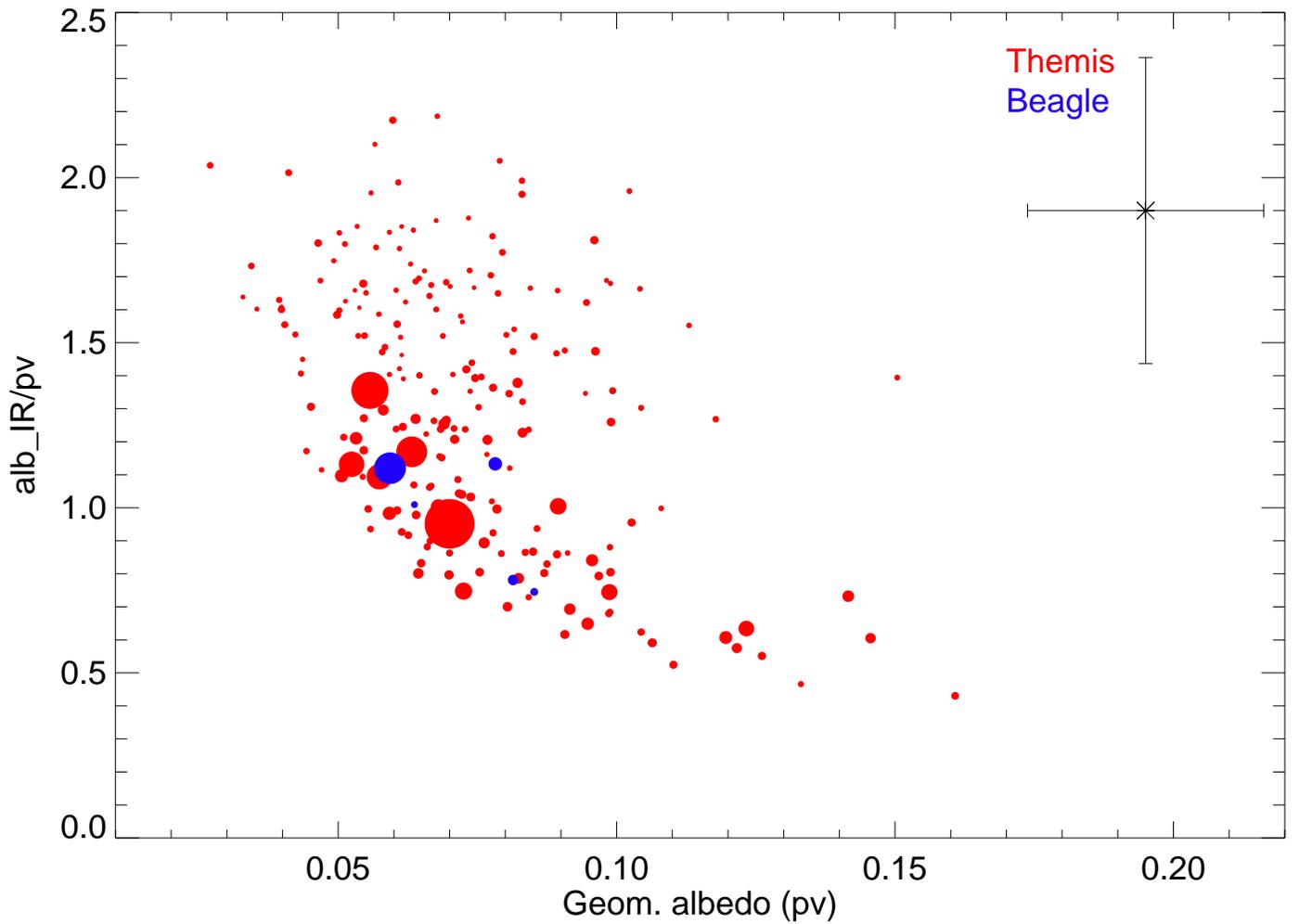,width=20truecm,angle=0}}
\caption{WISE albedos for the Themis family members: ratio of the infrared over the visible albedo (p$_v$) versus the visible albedo: Themis members having higher albedo value are also bluer in the IR range.
  The size of the symbols is proportional to the asteroids' diameters. In the upper rigth part of the plot we report the mean error bars of these measurements.}
\label{f10}
\end{figure*}

We also analyzed the near infrared albedo (p$_{IR}$), that is the albedo at 3.4--4.6 $\mu$m as defined in Mainzer et al. (2011a), and the geometric (p$_v$) albedo for the Themis members observed with WISE. Figure~\ref{f10} shows the ratio of the p$_{IR}$/p$_v$ albedo versus p$_v$ for a sample of 211 Themis and 5 Beagle members. The bright Themis members have lower p$_{IR}$/p$_v$ ratio, indicating a blue spectrum in the NIR region and/or the presence of absorption bands in the 3 $\mu$m region, potentially attributed to hydrated silicates, organics and eventually water ice.\\
A similar {\it waning moon} shape of the  p$_{IR}$/p$_v$ ratio versus p$_v$  was also noticed by Al\'i-Lagoa et al. (2013) analyzing B-type asteroids. Some biases may partially affect the distribution we see, in particular the lack of points in the lower left corner may be attributed to faint fluxes in WISE bands W1 and W2, not allowing a proper evaluation of the reflected sunlight contribution (Mainzer et al. 2011b).  Al\'i-Lagoa et al. (2013), in the analysis of B-type asteroids and p$_{IR}$ albedo, found that an absorption in the 3 $\mu$m region may be common on these bodies.  \\
Figure~\ref{f10} confirms the huge spectral variability of Themis members, which have small to moderate albedo values  (3--16\%) and a P$_{IR}$/p$_v$ ratio varying from $\sim$ 0.5 to 2.2.

\subsection{Meteorite analogues}

[Here Figs. 12-15 and Table 4]

To constrain the possible mineralogies of the investigated asteroids, we conducted
a search for meteorite and/or mineral spectral matches. We
sought matches only for objects observed across the entire VIS--NIR
wavelength range. We used the publicly available RELAB spectrum library (Pieters,
1983). For each  spectrum in the library, a filter was applied to find relevant  
wavelengths (0.4 to 2.45 $\mu$m). Then, a second filter was applied to reject spectra with irrelevant albedo values (i.e. brightness at 0.55 $\mu$m $>0.2$). A  Chi-squared value was calculated relative to the normalized input asteroid spectrum. We used the asteroid wavelength sampling to resample the laboratory spectra by linear interpolation. This allowed a least-squares calculation with the number of points equal to the wavelength sampling of the asteroid spectrum. The RELAB data files were sorted according to the corresponding Chi-squared values, and finally visually examined. A complete description of the search methodology
can be found in Fornasier et al. (2010, 2011).
 
\begin{figure*}
\centerline{\psfig{file=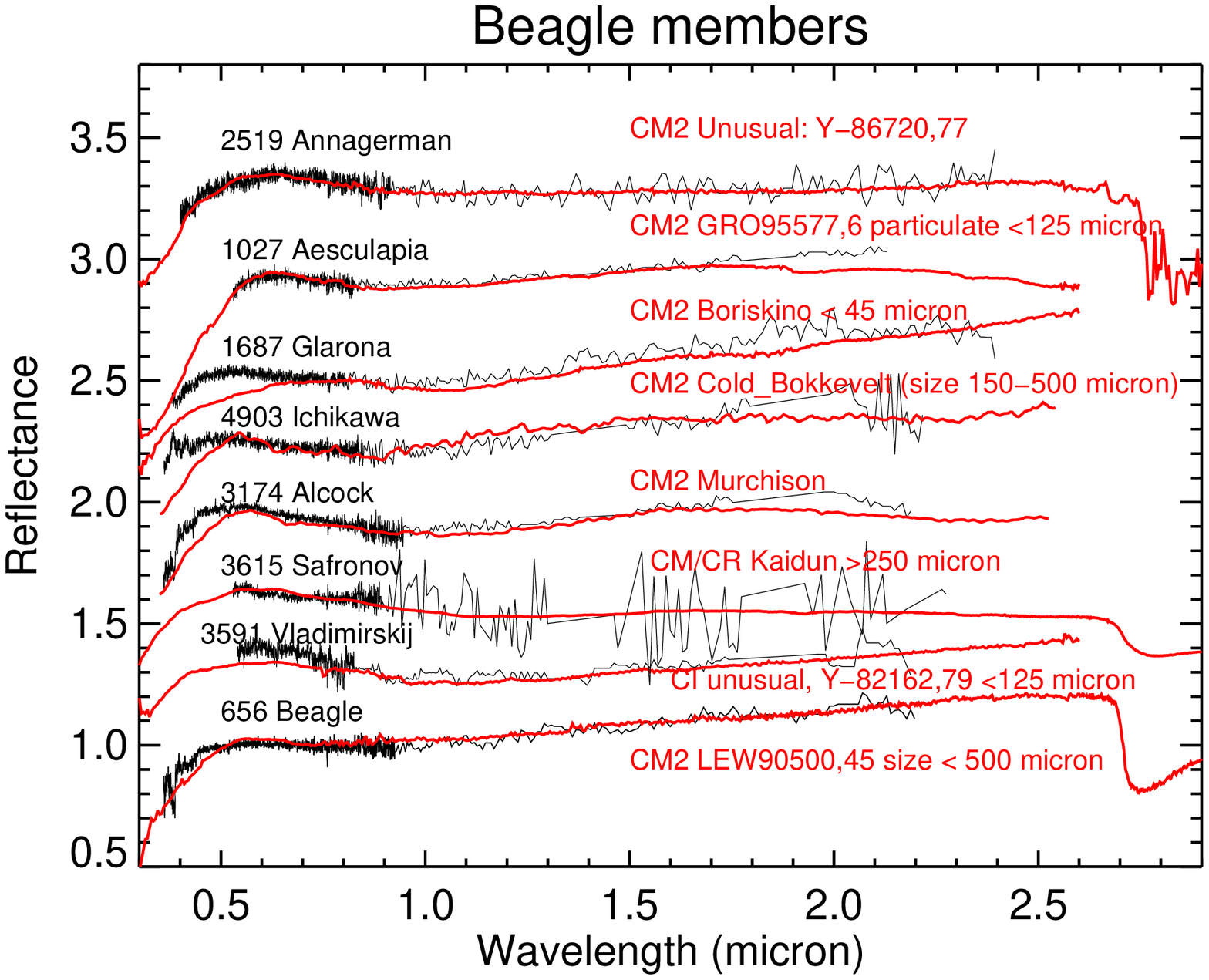,width=20truecm,angle=0}}
\caption{Best spectral matches between the observed Beagle family members and meteorites from the RELAB database.}
\label{f11}

\end{figure*}

\begin{figure}
\centerline{\psfig{file=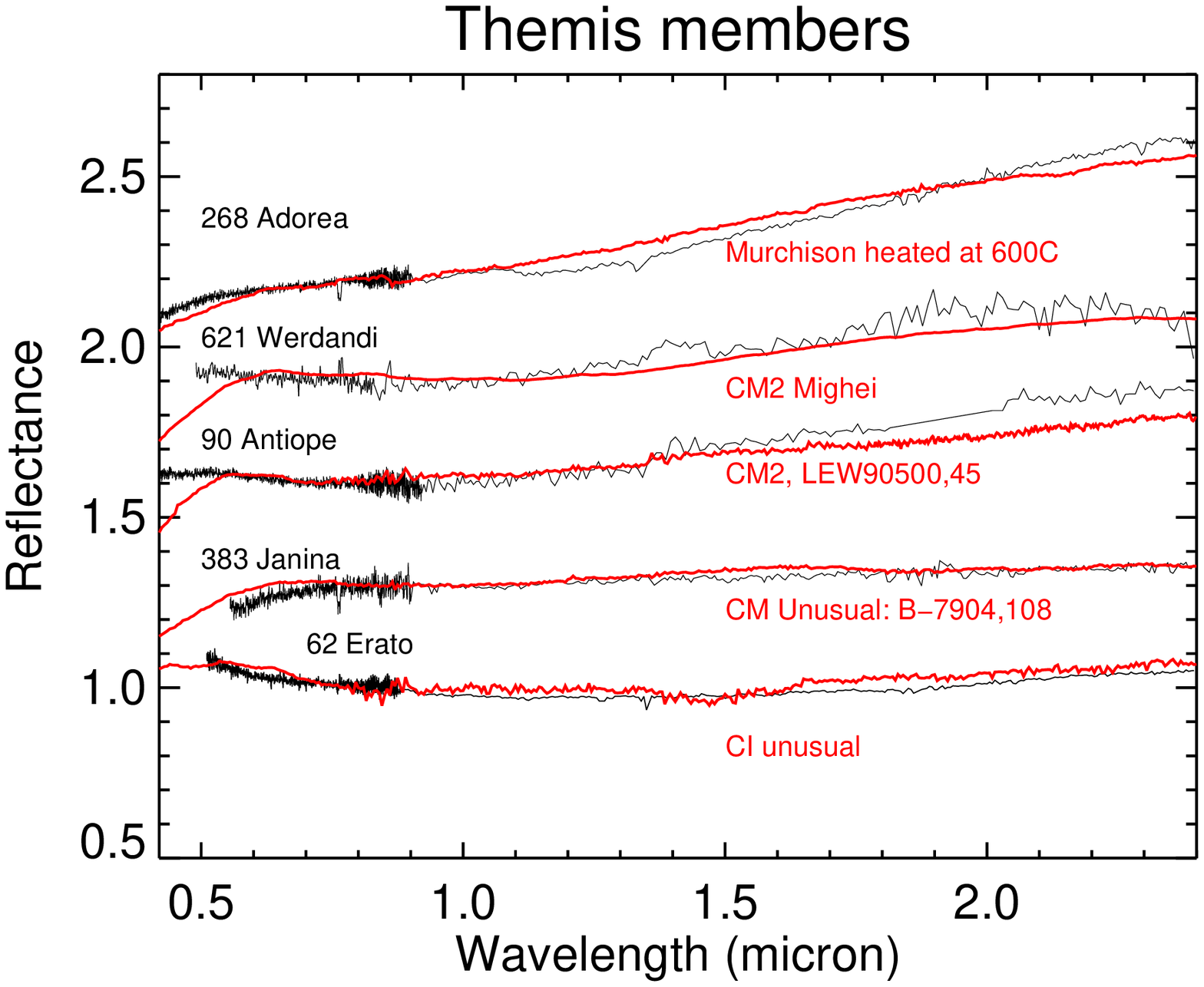,width=20truecm,angle=0}}
\caption{Best spectral matches between the observed Themis family members and meteorites from the RELAB database.}
\label{f12}
\end{figure}

\begin{figure}
\centerline{\psfig{file=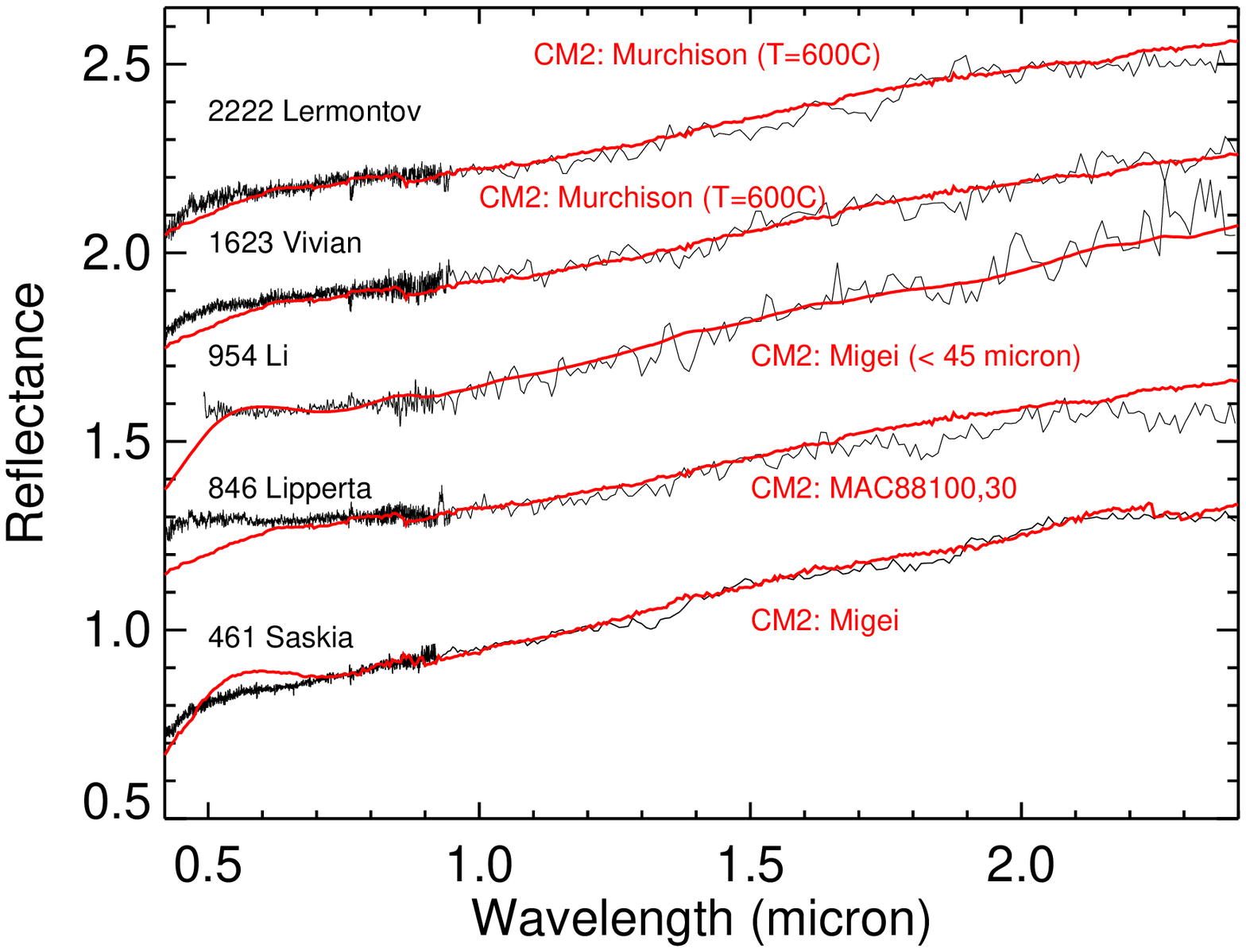,width=20truecm,angle=0}}
\caption{Best spectral matches between the observed Themis family members and meteorites from the RELAB database.}
\label{f13}
\end{figure}

\begin{figure}
\centerline{\psfig{file=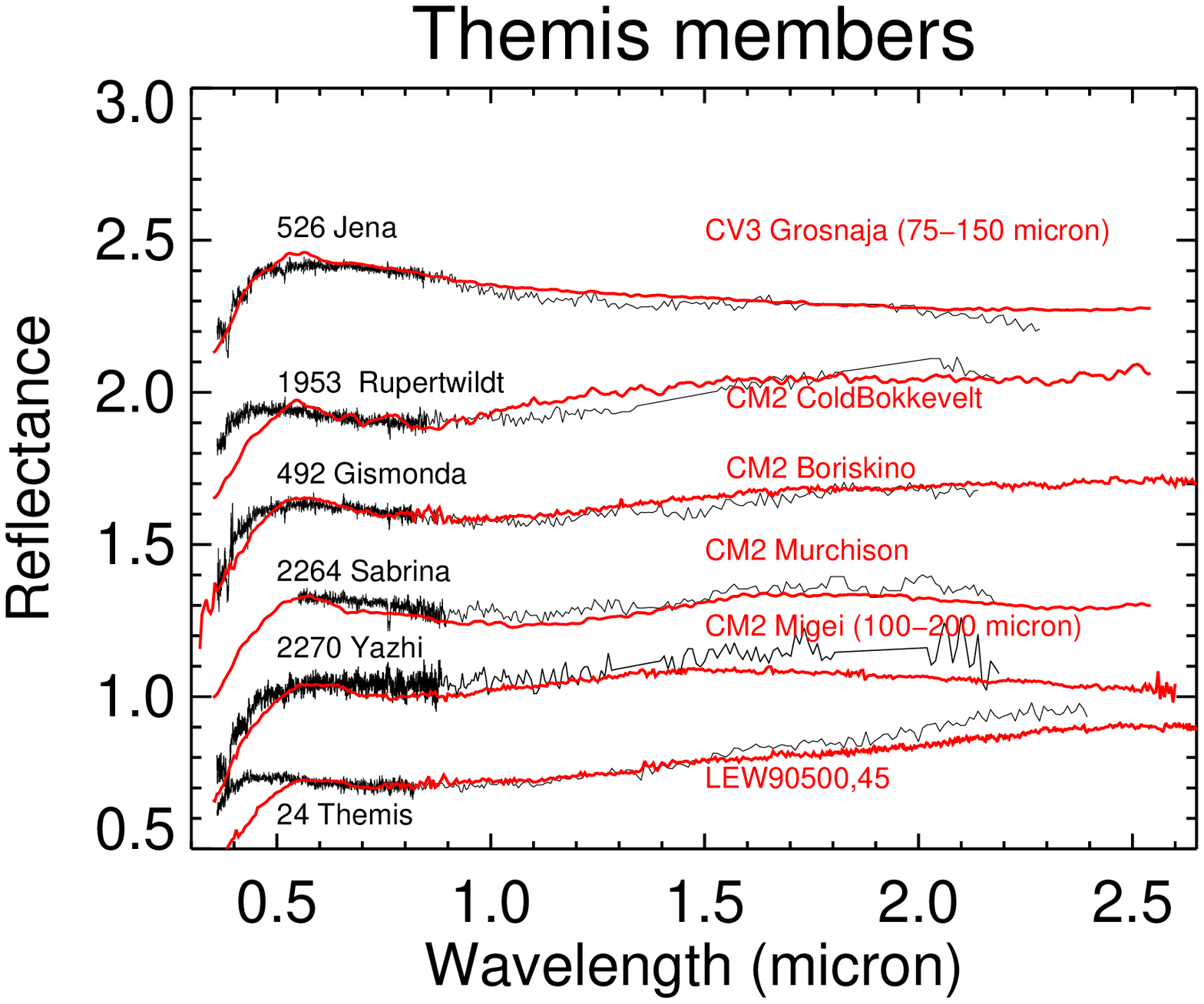,width=20truecm,angle=0}}
\caption{Best spectral matches between the observed Themis family members and meteorites from the RELAB database.}
\label{f14}
\end{figure}

The best matches between the observed asteroids and meteorites from the RELAB database are reported in Table~\ref{met}, and shown in Figures~\ref{f11} -~\ref{f14}.
Most of the Themis and Beagle families members are matched by CM2 carbonaceous chondrites, with few objects having CV3 or CI as best match. We do not see differences in the meteorite class matches between  the Beagle and Themis families members, except that there are no CV3 best match for Beagle members. Even if the aqueous alteration band at 0.7 $\mu$m has been identified just on one Beagle member (Kaluna et al., 2016), the meteorite matches indicate that these asteroids may have experienced aqueous alteration in the past. As previously stated, not all the aqueous altered  CM/CI meteorites show this feature (McAdam et al. 2015). \\
The asteroid (2519) Annagerman is spectrally perfectly matched by the unusual CI/CM Y-86720,77 with grain size  $<$ 125 $\mu$m, even if there are some differences in the albedo values, 11\% for the asteroid and  6\% for the meteorite. However it must be noted that the meteorite reflectance is usually evaluated at phase angle $>$ 7 $^o$ so it does not include the opposition surge effect. The composition and structure of the CM2 Y-86720 indicates that this meteorite was thoroughly aqueous altered (Lipschutz et al. 1999) and heated. Interestingly, some of the targets are best matched by heated or unusual meteorites. This was  already noticed by Clark et al. (2010), who found that Themis family asteroids tend to be best fitted by CM, CI, CR, CM/CI unusual meteorites, and/or heated CM/CI samples, and that 50\% of these asteroids are consistent with thermally metamorphosed material. \\
Our spectral matches are also in agreement with the de L\'eon et al. (2012) results from the  analysis of B-type asteroids, including 8 Themis members. These last fall in the G2-G5 groups as defined by de L\'eon et al. (2012), and are spectrally best matched by meteorites who experienced different stages of the aqueous alteration,  from the less altered CV3 meteorites up to the extensively altered CM2 chondrites. 
The spectral analysis together with the meteorite matches indicate that Themis members show a large spectral variety and different carbonaceous chondrite matches (from CV3 to CM2-CI), some of them being unusual or heated, indicating that these asteroids were thermally metamorphosed in the past.

\section{Discussion}

Spectroscopic studies of asteroid families can provide information about the interior of their parent bodies (e.g., Cellino et al. 2002); this can constrain models on the thermal and collisional evolution of asteroids.  More specifically, simulations of catastrophic disruption processes suggest that the ejected material coming from coherent sections of a heterogeneous progenitor body should maintain a coherent compositional structure (Michel et al. 2004). Therefore, compositional gradients within planetesimals should expose themselves within asteroid families (e.g., Jacobson et al. 2014).  To date, no remnants of a completely disrupted differentiated body have been identified. Vesta is the only family generated from a differentiated body (e.g., Lazzaro et al. 2004; Binzel and Xu 1993), but it is not a result of a catastrophic disruption, so its members come from the surface of the parent body.

The case of the Themis family seems to be unique. For example, Jacobson et al. (2014) used estimates of the progenitor mass and the mass of the largest remnant to assess the exposed nature of asteroid families, and he suggested that the Themis family is possibly the most exposed one.   We have found clear evidence of spectral diversity within the Themis family, and between Themis and Beagle families members.  There are several possible interpretations of this spectral diversity, which include a compositional gradient with depth in the original parent body (and/or with fragment size), space weathering effects, the survival of projectile fragments with a different composition, or a combination of these (e.g., Campins et al. 2012), that we discuss in the following. \\

\subsection{Scenario a): The Themis/Beagle family parent body was heterogeneous in composition} 

In this scenario, the diversity we see nowdays in Themis reflects different source regions in the parent body that was originally heterogeneous showing a compositional gradient. 
We clarify that a compositional gradient does not necessarily mean a metal core with a silicate mantle and crust. In fact, for primitive asteroids like the Themis parent body it was more likely a differentiation of rock and ice where the core underwent mild temperatures and no high thermal metamorphism. (e.g., Castillo-Rogez \& Schmidt 2010).
Consistent with this mild temperature model are the visible spectroscopic results of 36 members of the Themis family presented by Florczak et al. (1999). They found that about 50\% of their sample showed evidence of aqueous alteration, indicating that the parent body was sufficiently altered thermally to mobilize water. The results of  Florczak et al. (1999) also suggest that the percentage of asteroids showing aqueous alteration may decrease with the diameter of the objects, as expected by Vilas \& Sykes (1996) and proven by Rivkin (2012) and Fornasier et al. (2014) on a large sample of primitive main belt asteroids. The study  presented in this paper and in Fornasier et al. (2014) indicates a much lower percentage of hydration in the Themis family ($<$ 20\%), while just one out of the 45 Themis and Beagle small members studied by Kaluna et al. (2016) shows the 0.7 $\mu$m absorption band. Moreover the meteorite spectral matches found by us and in the literature point towards a similarity of Themis-Beagle family members with the CV3-CM2 chondrites, the meteorites having experienced aqueous alteration processes (see section 3.3; Clark et al. 2010; de L\'eon et al. 2012). \\
The above results support the view that at least part of the parent body of the Themis family was thermally altered and that the distribution of compositions could be attributed to fragments coming from different depths in the original parent body, which was catastrophically disrupted by a large projectile.\\
In this scenario, the Beagle family may have been originated by a fragment of a blue and bright piece of Themis parent body possibly coming from its interior.\\

\subsection{Scenario b): The Themis/Beagle family parent body was homogeneous and the projectile that collided with it was different in composition}

  Marzari et al. (1995) modeled the collisional evolution of the Themis family. Their best results yield parent body and projectile radii of 380 km and 190 km, respectively. They suggest that the two bodies were not formed in the same region and had different compositions. Since the projectile body was relatively large, its fragments should be a considerable part of the remnants. Interestingly, Florczak et al. (1999) argue that except for the hydration (indicated by an absorption centered near 0.7 $\mu$m) their visible spectra were homogeneous and they did not favor a different composition between parent body and projectile or significant differentiation.  Our visible and near-infrared spectra (Figs.~\ref{f7} and ~\ref{f8}) and those of de Le\'on et al. (2012) and Kaluna et al. (2016) show a wide and smooth range of behaviors suggestive of a range of compositions among the Themis family members. These results does not necessarily support different compositions between the large projectile and the Themis family parent body. Instead of a smooth spectral diversity, one might expect a bi-modal one from a different projectile and parent body composition.

\subsection{Scenario c): The family parent body was homogeneous and the spectral variability is produced by space weathering effects} 

  Assuming that the parent body of the Themis/Beagle families was quite homogeneous, the spectral variability seen on {\it old} Themis members and the fact the {\it young} Beagle members are bluer and with an higher albedo value than the Themis ones may be, at least in part, a result of space weathering effects. 

In the past 20 years enormous progress has been reached in the study of space weathering (SW) effects on silicates and S-type asteroids. The so-called ordinary chondrite paradox, that is lack of asteroids similar to the ordinary chondrites, which represent 80\% of meteorite falls, has been solved. These meteorites are now clearly related to S-type asteroids, as proved  by direct measurements of the NEAR and HAYABUSA missions on the near Earth asteroids Eros and Itokawa (Clark et al. 2001; Noguchi et al. 2011) and laboratory experiments on irradiation of silicates/ordinary chondrites (Moroz et al. 1996; Sasaki et al. 2002; Brunetto \& Strazzulla 2005; Noble et al. 2007). Spectral differences between S-type asteroids and ordinary chondrites are caused by space weathering effects, which produce a darkening in the albedo, a reddening of the spectra, and diminish the silicates absorption bands on the asteroids surfaces, exposed to cosmic radiation and solar wind. \\
On the other hand, our understanding of space weathering effects on primitive asteroids is still poor.
Only few laboratory experiments have been devoted to the investigation of SW effects on low albedo carbonaceous chondrites and on complex organic materials, and they indicate a no linear trend but much more complex effects. 
Irradiation of transparent organic materials produces  firstly  redder and lower albedo materials  that upon  further processing  turn  flatter-bluishing and darker (Kanuchova et  al. 2012; Moroz et al. 2004a). This result is consistent with a recent study that was looking for spectral differences between Ch/Cgh-types and CM chondrites (Lantz et al. 2013). On the other hand, Lazzarin et al. (2006) expected a global reddening for the whole asteroidal population. Laboratory experiments on carbonaceous chondrites  give different results, that is reddening and/or blueing. Some ion irradiations on CV Allende and CO Frontier Mountain 95002 (Lazzarin et al. 2006) led to reddening and darkening, and this trend  is confirmed for the Allende meteorite using both ion (Brunetto et al. 2014) and laser irradiation (Gillis-Davis et al. 2015).  \\
The laser irradiated CM Mighei resulted in spectral reddening (Moroz et al. 2004b) while laser irradiation on Tagish Lake showed a blueing effect (Hiroi et al. 2004).  Vernazza et al. (2013) performed ion irradiations on Tagish Lake and observed a strong blueing with argon, while irradiation with He produced only minor spectral changes and a slight reddening at 4 keV dose. Hiroi et al. (2013) used laser irradiations on a CO, a CK, a CM, a CI and Tagish Lake: the first two present the same behavior as ordinary chondrites while the others show a clear blueing (but reflectance decreases so that a darkening effect is seen). \\
For the CM Murchison the results after irradiation are contradictory:  Keller et al. (2015) found reddening and darkening of the sample after ion irradiation,  Matsuoka et al. (2015) found a blueing and darkening effect with laser irradiation,  Gillis-Davis et al. (2015), using again laser irradiation, found a darkening effect but without slope variations, while Lantz et al. (2015) found no clear albedo and slope variations after ion irradiation.  

It appears complicated to define a general space weathering trend for the primitive objects when considering the various results of laboratory simulations on carbonaceous chondrites. That would explain why the few studies on asteroids
proposed different conclusions between reddening (Lazzarin et al. 2006) and blueing of the slope (Nesvorny et al. 2005; Lantz et al. 2013).
Here we compare spectra of young and old members of the Themis group (respectively Beagle and Themis families). Starting from the hypothesis that the parent body was homogeneous, the Beagle members have been less exposed to the harsh space environment, and so less space weathered compared to Themis members. If this hypothesis is correct, then our observations seem to indicate that the surfaces of the Themis family members have the same spectral answer than the S-type asteroids: young Beagle members have globally higher albedo and bluer slopes, while most of the older Themis members are redder and have a lower albedo value. A similar conclusion is found by Kaluna et al. (2016) on their analysis of Themis and Beagle members of similar size, smaller than 15 km. From the mean slopes values of the two families, Kaluna et al. (2016) estimate a slope reddening of 0.08 $\mu$m$^{-1}$ in 2.3 Gyr for C-complex asteroids. However, if space weathering effects are the only cause of the spectral diversity between the two families, we would expect a distinct spectral and albedo distribution between old and young asteroids, while our analysis clearly shows that several Themis members have albedo and spectral properties similar to the Beagle ones (Figure~\ref{f8}). Some rejuvenating processes might have taken place (Shestopalov et al. 2013), but they cannot solely explain the large percentage of blue and bright Themis members observed. \\
Another important point is the timescale needed to space weathering processes to alter asteroid surfaces. Laboratory simulations indicate that solar wind may alter the surfaces in a relatively short timescale, of the order of 10$^3$--10$^5$ years, producing darkening and reddening effects on ordinary chondrites (Strazzulla et al. 2005), on S-type asteroids (Vernazza et al. 2009a, 2009b), and on the Allende carbonaceous chondrite (Brunetto et al. 2014). This has also been confirmed by the Itokawa grain analysis brought back to the Earth by the sample return mission Hayabusa (Noguchi et al. 2011). If these short timescales turn out to be correct, it is possible that Beagle members are old enought to have already experienced some space weathering processes. All this considered, we cannot firmly conclude that the observed spectral diversity is related to space weathering processes. Considering the presence of hydrated minerals on some members and the thermal evolution models which suggest a differentiation in the Themis/Beagle parent body, we favor the compositional heterogeneity in the original parent body as explanation of the spectral diversity within Themis/Beagle members, even if we cannot exclude that space weathering processes may be responsible of at least part of the spectral differences observed.

\section{Conclusions}

In this work we investigate the spectral properties of primitive asteroids belonging to the $\sim$ 2 Gyr old Themis family and to the young ($<$ 10 Myr) Beagle subcluster within the Themis family.  We present new visible and near infrared spectra of 22 Themis and 8 Beagle families members. The Themis family members show different spectral behaviors including asteroids 
with blue/neutral and moderately red spectra, while the Beagle members have less spectral variability and they are all neutral/blue.
We include in our analysis most of the data available in the literature on Themis/Beagle families (a sample of 79 objects) for a complete analysis of their spectral properties versus physical parameters. The spectral slope versus albedo distribution clearly shows that Beagle members are less red and tend to have an higher albedo value than the Themis one. To explain the spectral differences between the two families we propose different interpretations, including heterogeneity in the parent body having a compositional gradient with depth, the survival of projectile fragments having a different composition than the parent body, space weathering effects, or a combination of these. \\
The spectral differences observed between Beagle and Themis members may be partially attributed to space weathering effects, which would produce on primitive asteroids reddening and darkening, as seen on S-type asteroids. However, space weathering effects cannot solely explain the spectral variety within the two families, first because we observe several Themis members having albedo and spectral behaviors similar to the younger Beagle members, and second because we need to assume that the parent body of the family was homogeneous. However, our and literature data show the presence of hydrated minerals on some of the Themis members, and a spectral analogy of the Themis/Beagle members with carbonaceous chondrites having experienced different degrees of aqueous alteration. These results, together with those from thermal evolution models, indicate that the parent body of the Themis family experienced mild thermal metamorphism in the past, with a possible compositional gradient with depth. We thus favor the compositional heterogeneity in the original parent body as main source of the spectral diversity within Themis/Beagle members.

\vspace{0.3truecm}
{\bf Acknowledgment} \\
This paper is based on observations made with the Italian Telescopio Nazionale Galileo (TNG) operated on the island of La Palma by the Centro Galileo Galilei of the INAF (Istituto Nazionale di
Astrofisica), both located at the Observatorio del Roque de los Muchachos, La Palma, Spain, of the Instituto de Astrofisica de Canarias." This project was supported by the French Planetology National Programme (INSU-PNP). The authors gratefully acknowledge the reviewers for their comments and suggestions who help us improving the paper.

\bigskip

{\bf References} \\

Al\'i-Lagoa, V., de Le\'on, J., Licandro, J., Delb\`o, M., Campins, H., Pinilla-Alonso, N., Kelley, M. S., 2013. Physical properties of B-type asteroids from WISE data. Astron. Astroph. 554, id. A71, 16 pp

Baffa, C. et al., 2001. NICS: The TNG Near Infrared Camera Spectrometer. Astron. Astrophys. 378, 722--728

Beck,P., Quirico, E.,  Sevestre, D., Montes-Hernandez,D.,  Pommerol A., Schmitt, B., 2011. Goethite as an alternative origin of the 3.1 micron band on dark asteroids. Astron. Astroph.  526, id. A85, 4pp

Binzel, R., and Xu, S., 1993.  Chips off of asteroid 4 Vesta - Evidence for the parent body of basaltic achondrite meteorites.  Science 260, 186--191
	
Bishop, J.L., \& Pieters, C. 1995. Low-temperature and low atmospheric pressure infrared reflectance spectroscopy of Mars soil analog materials.
J. of Geoph. Res. 100, 5369-5379

Bro\v{z}, M., Morbidelli, A., Bottke, W. F., Rozehnal, J., Vokrouhlicky, D., Nesvorny, D.  2013. Constraining the cometary flux through the asteroid belt during the late heavy bombardment. Astron. Astroph. id. A117, 16 pp

Brunetto, R., Strazzulla, G., 2005. Elastic collisions in ion irradiation experiments: A mechanism for space weathering of silicates. Icarus 179, 265-273

Brunetto, R., Lantz, C., Ledu, D., Baklouti, D., Barucci, M.A., Beck, P., Delauche, L., Dionnet, Z., Dumas, P., Duprat, J., Engrand, C., Jamme, F., Oudayer, P., Quirico, E., Sandt, C., Dartois, E., 2014. Ion irradiation of Allende meteorite probed by visible, IR, and Raman spectroscopies. Icarus 237, 278-292

Bus, S. J., Binzel, R.P., 2002. Phase II of the Small Main-Belt Asteroid Spectroscopic Survey. The Observations. Icarus 158, 106--145

Bus, S. and Binzel, R. P., 2003. Small Main-belt Asteroid Spectroscopic Survey, Phase II. EAR-A-I0028-4-SBN0001/SMASSII-V1.0. NASA Planetary Data System.

Campins, H., Hargrove, K.,  Pinilla-Alonso,N., Howell, E.S., Kelley,  M.S.,  Licandro, J., Moth\'e-Diniz, T.,  Fernandez, Y.,  Ziffer, J., 2010. Water ice and organics on the surface of the asteroid 24 Themis. Nature 464, 1320--1321

Campins, H. de Le\'on, J., Licandro, J., Kelley, M. S., Fernandez, Y., Ziffer, J., Nesvorny, D., 2012. Spectra of asteroid families in support of Gaia. Plan. and Space Sci. 73, 95--97

Castillo-Rogez, J.C., Schmidt, B.E., 2010. Geophysical evolution of the Themis family
parent body. Geophys. Res. Lett. 37, L10202

Cellino, A., Bus, S. J., Doressoundiram, A., Lazzaro, D., 2002. Spectroscopic Properties of Asteroid Families. In: Bottke, W.F., Jr., Cellino, A., Paolicchi, P.,
Binzel, R.P. (Eds.), Asteroids, vol. III. Univ. of Arizona Press, Tucson, pp. 633--643

Clark, B.E., Lucey, P., Helfenstein, P., Bell, III, J.F., Peterson, C., Veverka, J., McConnochie, T., Robinson, M.S., Bussey, B., Murchie, S.L., Izenberg, N.I., Chapman, C.R., 2001. Space weathering on Eros: Constraints from albedo and spectral measurements of Psyche crater. Meteorit. Planet. Sci. 36, 1617-1637
	
Clark, B. E., Ziffer, J. Nesvorny, D., Campins, H., Rivkin, A. S., Hiroi, T. Barucci, M.A., Fulchignoni, M. Binzel, R. P., Fornasier, S. DeMeo, F. Ockert-Bell, M. E., Licandro, J. Moth\`e-Diniz, T., 2010. Spectroscopy of B-type asteroids: Subgroups and meteorite analogs. J. of Geoph. Res. 115, ID E06005 

De L\'eon, J., Pinilla-Alonso, N., Campins, H., Licandro, J., Marzo, G. A., 2012. Near-infrared spectroscopic survey of B-type asteroids:
Compositional analysis. Icarus 218, 196--206

Durda, D. D., Bottke Jr., W. F., Nesvorny, D., Enke, B. L., Merline, W. J., Asphaug, E., Richardson, D. C., 2007. Size-frequency distributions of fragments from SPH/ N-body simulations of asteroid impacts: Comparison with observed asteroid families. Icarus 186, 498--516

Fornasier, S., Lazzarin, M., Barbieri, C., Barucci, M. A., 1999. Spectroscopic comparison of aqueous altered asteroids with CM2 carbonaceous chondrite meteorites.  Astron. Astrophys. 135, 65--73  

Fornasier, S., Dotto, E., Barucci, M. A.,  Barbieri, C. 2004. Water ice on the surface of the large TNO 2004 DW. Astron. Astrophys. 422, 43--46

Fornasier, S., Migliorini, A., Dotto, E., Barucci, M.A., 2008. Visible and near infrared
spectroscopic investigation of E-type asteroids, including 2867 Steins, a target
of the Rosetta mission. Icarus 196, 119--134

Fornasier, S., Clark, B.E., Dotto, E., Migliorini, Ockert-Bell, M., Barucci, M.A., 2010. Spectroscopic survey of M-type asteroids. Icarus 210, 655--673.

Fornasier, S., Clark, B. E., Dotto, E., 2011. Spectroscopic survey of X-type asteroids. Icarus 214, 131--146

Fornasier, S., Lantz, C, Barucci, M.A., Lazzarin, M., 2014. Aqueous alteration on Main Belt primitive asteroids:
Results from visible spectroscopy. Icarus 233, 163--178

Florczak M., Lazzaro, D.,  Moth\'{e}--Diniz, Angeli, CA,  Betzler A.S., 1999. A spectroscopic study of the Themis family. Astron. Astrophys. Suppl. Ser. 134, 463--471 

Gillis-Davis, J.J., Gasda, P.J., Bradley, J.P., Ishii, H.A., Bussey, D.B.J., 2015. Laser Space Weathering of Allende (CV2) and Murchison (CM2) Carbonaceous Chondrites. Lunar and Planetary Science Conference 46, 1607

Haghighipour, N., 2009. Dynamical constraints on the origin of main belt comets. Meteorit. Planet. Sci. 44, 1863--1869

Hargrove, K. D., Emery, J. P., Campins, H., Kelley, M. S. P., 2015. Asteroid (90) Antiope: Another icy member of the Themis family? Icarus 254, 150--156

Higley, S., Hardersen, P., Dyvig, R. 2008. Shape and Spin Axis Models for 2 Pallas (Revisited) 5 Astraea, 24 Themis, and 105 Artemis. The Minor Planet Bulletin 35, 63--66

Hirayama K., 1918. Groups of asteroids probably of common origin.  Astron. J. 31 185--188

Hiroi, T., Pieters, C.M., Rutherford, M.J., Zolensky, M.E., Sasaki, S., Ueda, Y., Miyamoto, M., 2004. What are the P-type Asteroids Made Of?. Lunar and Planetary Science Conference 35, 1616

Hiroi, T., Sasaki, S.,  Misu, T., Nakamura, T., 2013. Keys to Detect Space Weathering on Vesta: Changes of Visible and Near-Infrared Reflectance Spectra of HEDs and Carbonaceous Chondrites. Lunar and Planetary Science Conference 44, 1276

Hsieh, H. H., \& Jewitt, D., 2006. A Population of Comets in the Main Asteroid Belt. Science 312, 561--563

Hsieh, H. H., Jewitt, D., Fernandez, Y. R., 2009. Albedos of Main-Belt Comets 133P/Elst-Pizarro and 176P/LINEAR. Astroph. J. Letters 694,  L111-L114

Hsieh, H. H., Yang, B., Haghighipour, N., Kaluna, H. M., Fitzsimmons, A., Denneau, L., Novakovic, B., Jedicke, R., Wainscoat, R.J., Armstrong, J. D., and 32 coauthors, 2012. Discovery of Main-belt Comet P/2006 VW139 by Pan-STARRS1. Astroph. J. Letters 748, L15, 7 pp

Jacobson, S., Campins, H., Delb\'o, M., Michel, P., Tanga, P., Hanus, J., Morbidelli, A., 2014. Using asteroid families to test planetesimal differentiation hypotheses.
Asteroids, Comets, Meteors 2014. Proceedings of the conference held 30 June -- 4 July, 2014 in Helsinki, Finland. Edited by K. Muinonen et al.

Jewitt, D, \& Guilbert-Lepoutre, A., 2012. Limits to Ice on Asteroids (24) Themis and (65) Cybele. Astron. J. 143, 21, 8pp

Jones, T. D., Lebofsky, L. A., Lewis, J. S.,  Marley, M. S., 1990. The composition and origin of the C, P, and D asteroids: Water
as a tracer of thermal evolution in the outer belt. Icarus 88, 172--192

Kaluna, H. M., Masiero, J, R., Meech, K, J, 2016. Space Weathering Trends Among Carbonaceous Asteroids. Icarus 264, 62--71

Ka{\v n}uchov{\'a}, Z., Brunetto, R., Melita, M., Strazzulla, G., 2012. Space weathering and the color indexes of minor bodies in the outer Solar System. Icarus 221, 12-19

Keller, L.P., Christoffersen, R., Dukes, C.A., Baragiola, R., Rahman, Z., 2015. Ion Irradiation Experiments on the Murchison CM2 Carbonaceous Chondrite: Simulating Space Weathering of Primitive Asteroids. Lunar and Plan. Sci. Conference 46, 1913

Lantz, C., Clark, B.E., Barucci, M.A., Lauretta, D.S., 2013. Evidence for the effects of space weathering spectral signatures on low albedo asteroids. Astron. Astroph. 554, id. A138, 7 pp

Lantz, C., Brunetto, R., Barucci, M.A., Dartois, E., Duprat, J., Engrand, C., Godard, M., Ledu, D., Quirico, E., 2015. Ion irradiation of the Murchison meteorite: Visible to mid-infrared spectroscopic results. Astron. Astrophys. 577, id. A41, 9 pp

Lazzarin, M., Marchi, S., Moroz, L.V., Brunetto, R., Magrin, S., Paolicchi, P., Strazzulla, G., 2006. Space Weathering in the Main Asteroid Belt: The Big Picture. The Astroph. J. Letters 647, L179-L182

Lazzaro, D., Angeli, C. A., Carvano, J. M., Moth\'e-Diniz, T., Duffard, R., Florczak, M., 2004. S$^{3}$OS$^{2}$: the visible spectroscopic survey of 820 asteroids. Icarus 172, 179--220

Lebofsky L.A. 1980. Infrared reflectance spectra of asteroids: A search for water of hydration. Astron. J. 85, 573--585

Licandro, J., Campins, H., Tozzi, G. P., de Le\'on, J., Pinilla-Alonso, N., Boehnhardt, H., Hainaut, O. R., 2011. Testing the comet nature of main belt comets. The spectra of 133P/Elst-Pizarro and 176P/LINEAR.  Astron. Astrophys. 532, id. A65, 7pp. 

Licandro, J., Hargrove, K., Kelley, M., Campins, H., Ziffer, J., Al\'i-Lagoa, V.., Fernandez, Y., Rivkin, A., 2012. 5-14 micron  Spitzer spectra of Themis family asteroids. Astron. Astrophys. 537, id. A73, 7 pp

Licandro, J., Moreno, F., de Le\'on, J., Tozzi, G. P., Lara, L. M., Cabrera-Lavers, A., 2013. Exploring the nature of new main-belt comets with the 10.4 m GTC telescope: (300163) 2006 VW139.  Astron. Astrophys. 550, id. A17, 7pp.

Lipschutz, M.E., Zolensky M.E., Bell, M.S., 1999. New petrographic and trace element data on thermally metamorphosed carbonaceous chondrites. Antarct. Met. Res. 12, 57--80

Lunine, 2006, Meteorites and the Early Solar System II, Univ. of Arizona Press, Tucson, pp. 309--319

Mainzer, A., Grav, T., Masiero, J., Bauer, J., Wright, E., Cutri, R. M., McMillan, R. S., Cohen, M., Ressler, M., Eisenhardt, P., 2011a. Thermal Model Calibration for Minor Planets Observed with Wide-field Infrared Survey Explorer/NEOWISE. The Astroph. J. Letters 736, id. 100, 9 pp

Mainzer, A., Grav, T., Masiero, J., Hand, E., Bauer, J., Tholen, D., McMillan, R. S., Spahr, T., Cutri, R. M., Wright, E., Watkins, J., Mo, W., Maleszewski, C., 2011b. NEOWISE Studies of Spectrophotometrically Classified Asteroids: Preliminary Results. Astroph. J. 741, id. 90, 25 pp

Marzari, F.,  Davis D., Vanzani, V. 1995. Collisional evolution of asteroid families. Icarus 113, 168--187

Masiero, J.R., Mainzer, A.K., Grav, T., Bauer, J.M., Cutri, R.M., Dailey, J., Eisenhardt, P.R.M., McMillan, R.S., Spahr, T.B., Skrutskie, M.F., Tholen, D., Walker, R.G.,
Wright, E.L., DeBaun, E., Elsbury, D., Gautier, T., Gomillion, S., Wilkins, A., 2011. Main belt asteroids with WISE/NEOWISE. I. Preliminary albedos and diameters. Astrophys. J. 741, 68, 1--20

Matsuoka, M., Nakamura, T., Kimura, Y., Hiroi, T., Nakamura, R., Okumura, S., Sasaki, S., 2015. Pulse-laser irradiation experiments of Murchison CM2 chondrite for reproducing space weathering on C-type asteroids. Icarus 254, 135--143
	
McAdam, M. M., Sunshine, J. M., Howard, K. T., McCoy, T. M., 2015. Aqueous alteration on asteroids: Linking the mineralogy and spectroscopy of CM and CI chondrites. Icarus 245, 320--332

Michel, P., Benz, Willy, Richardson, Derek C., 2004. Catastrophic disruption of pre-shattered parent bodies. Icarus 168, 420--432

Moth\'{e}-Diniz, T., Roig, F., Carvano, J.M., 2005. Reanalysis of asteroid families structure through visible spectroscopy. Icarus 174, 54--80.

Morbidelli, A, Chambers, J., Lunine, J. I., Petit, J. M., Robert, F., Valsecchi, G. B., Cyr, K. E., 2000. Source regions and time scales for the delivery of water to Earth. Meteor. \& Plan. Sci. 35, 1309--1320

Moroz, L.V., Fisenko, A.V., Semjonova, L.F., Pieters, C.M., Korotaeva, N.N., 1996. Optical Effects of Regolith Processes on S-Asteroids as Simulated by Laser Shots on Ordinary Chondrite and Other Mafic Materials. Icarus 122, 366--382

Moroz, L., Baratta, G., Strazzulla, G., Starukhina, L., Dotto, E., Barucci, M.A., Arnold, G., Distefano, E., 2004a. Optical alteration of complex organics induced by ion irradiation:. 1. Laboratory experiments suggest unusual space weathering trend. Icarus 170, 214--228

Moroz, L.V., Hiroi, T., Shingareva, T.V., Basilevsky, A.T., Fisenko, A.V., Semjonova, L.F., Pieters, C.M., 2004b. Reflectance Spectra of CM2 Chondrite Mighei Irradiated with Pulsed Laser and Implications for Low-Albedo Asteroids and Martian Moons. Lunar and Planetary Science Conference 35, 1279

Nesvorny, D., Jedicke, R., Whiteley, R.J., Ivezic, Z., 2005. Evidence for asteroid space weathering from the Sloan Digital Sky Survey. Icarus 173, 132--152

Nesvorny, D., W.F. Bottke, D. Vokrouhlicky, M. Sykes, D.J. Lien, J. Stansberry, 2008. Origin of the Near-Ecliptic Circumsolar Dust Band. Astrophys. J. 679, L143--L146

Nesvorny, D., Nesvorny HCM Asteroid Families V2.0. EAR-A-VARGBDET-5-NESVORNYFAM-V2.0. NASA Planetary Data System, 2012

Noble, S.K., Pieters, C.M., Keller, L.P., 2007. An experimental approach to understanding the optical effects of space weathering. Icarus 192, 629--642

Noguchi, T., Nakamura, T., Kimura, M., Zolensky, M.E., Tanaka, M., Hashimoto, T., Konno, M., Nakato, A., Ogami, T., Fujimura, A., Abe, M., Yada, T., Mukai, T., Ueno, M., Okada, T., Shirai, K., Ishibashi, Y., Okazaki, R., 2011. Incipient Space Weathering Observed on the Surface of Itokawa Dust Particles. Science 333, 1121--1125

Perna, D., Kanuchov\'a, Z., Ieva, S., Fornasier, S., Barucci, M. A., Lantz, C., Dotto, E., Strazzulla, G., 2015. Short-term variability on the surface of (1) Ceres⋆. A changing amount of water ice? Astron. Astroph. 575, Id. L1, 6 pp

Pieters, C., 1983. Strength of mineral absorption features in the transmitted component of near-infrared reflected light: First results from RELAB. J. Geophys.
Res. 88, 9534--9544

Reddy, V., Nathues, A., Le Corre, L., Li, J.-Y., Schafer, M., Hoffmann, M., Russel,  C. T., Mengel, K., Sierks, H., Christensen, U., 2015.Nature of Bright Spots on Ceres from the Dawn Framing Camera.  LPI Contribution No. 1856, p. 5161

Rivkin, A.S., Emery, J.P., 2010. Detection of ice and organics on an asteroidal surface. Nature 64, 1322--1323

Rivkin, A.S., 2012. The fraction of hydrated C-complex asteroids in the asteroid belt from SDSS data. Icarus 221, 744--752

Sasaki, S., Hiroi, T., Nakamura, K., Hamabe, Y., Kurahashi, E., Yamada, M., 2002. Simulation of space weathering by nanosecond pulse laser heating: dependence on mineral composition, weathering trend of asteroids and discovery of nanophase iron particles. Advances in Space Research 29, 783--788

Shestopalov, D.I., Golubeva, L.F., Cloutis, E.A., 2013. Optical maturation of asteroid surfaces. Icarus 225, 781--793.

Strazzulla, G., Dotto, E., Binzel, R., Brunetto, R., Barucci, M.A., Blanco, A., Orofino, V., 2005. Spectral alteration of the Meteorite Epinal (H5) induced by heavy ion
irradiation: A simulation of space weathering effects on near-Earth asteroids. Icarus 174, 31--35

Tholen, D.J., 1984. Asteroid taxonomy from cluster analysis of photometry. Ph.D. dissertation, University of Arizona, Tucson.

Usui, F., Kuroda, D., M\"uller, T. G., Hasegawa, S., Ishiguro, M., Ootsubo, T., Ishihara, D., Kataza, H., Takita, S., Oyabu, S., Ueno, M., Matsuhara, H., Onaka, T., 2011. Asteroid Catalog Using Akari: AKARI/IRC Mid-Infrared Asteroid Survey. Astronom. Soc. of Japan 63, 1117--1138

Vernazza, P., Binzel, R.P., Rossi, A., Fulchignoni, M., Birlan, M., 2009a. Solar wind as the origin of rapid reddening of asteroid surfaces. Nature 458, 993--995

Vernazza, P., Brunetto, R., Binzel, R.P., Perron, C., Fulvio, D., Strazzulla, G., Fulchignoni, M., 2009b. Plausible parent bodies for enstatite chondrites and mesosiderites: Implications for Lutetia’s fly-by. Icarus 202, 477--486

Vernazza, P., Fulvio, D., Brunetto, R., Emery, J.P., Dukes, C.A., Cipriani, F., Witasse, O., Schaible, M.J., Zanda, B., Strazzulla, G., Baragiola, R.A., 2013. Paucity of Tagish Lake-like parent bodies in the Asteroid Belt and among Jupiter Trojans. Icarus 225, 517--525

Vilas, F.,\& Gaffey, M.J., 1989. Phyllosilicate absorption features in Main-Belt and Outer-Belt asteroid reflectance spectra. Science 246, 790--792

Vilas, F., Hatch, E.C., Larson, S.M., Sawyer, S.R., Gaffey, M.J., 1993. Ferric iron in primitive asteroids - A 0.43-$\mu$m absorption feature. Icarus 102, 225--231

Vilas, F., 1994. A cheaper, faster, better way to detect water of hydration on Solar System bodies. Icarus 111, 456--467

Vilas, F., Jarvis, K.S., Gaffey, M.J., 1994. Iron alteration minerals in the visible and near-infrared spectra of low-albedo asteroids. Icarus 109,
274--283

Vilas, F., \& Sykes M. W., 1996. Are Low-Albedo Asteroids Thermally Metamorphosed? Icarus 124, 483--489

Zappala, V., Cellino, A., Farinella, P., Knezevic, Z., 1990. Asteroid families. I. Identification by hierarchical clustering and reliability assessment. Astron. J. 100, 2030--2046

Zappala, V., Bendjoya, Ph., Cellino, A., Farinella, P., Froeschl\'e, C.,  1995. Asteroid families: Search of a 12,487-asteroid sample using
two different clustering techniques. Icarus 116, 291--314

Ziffer, J., Campins, H., Licandro, J., Walker, M. E., Fernandez, Y. Clark, B. E., Mothe-Diniz, T., Howell, E. Deshpande, R., 2011. Near-infrared spectroscopy of primitive asteroid families.
Icarus 213, 538--546

Xu, S. Binzel, R. P., Burbine, T. H., Bus, S. J. 1995. Small main-belt asteroid spectroscopic survey: Initial results. Icarus 155, 1--35


\newpage


         {\bf Figures}

\newpage

{\bf Tables}

{\scriptsize
       \begin{center}
     \begin{longtable} {|l|l|c|c|c|c|c|c|l|}  
\caption[]{Observational circumstances. Solar analog stars named "HIP"
  come from the Hipparcos catalogue and "Lan" from the
Landolt photometric standard stars catalogue.}. 
        \label{tab1} \\
\hline \multicolumn{1}{|c|} {\textbf{Object    }} & \multicolumn{1}{c|}
{\textbf{Night}} & \multicolumn{1}{c|} {\textbf{UT$_{start}$}} &
\multicolumn{1}{c|} {\textbf{T$_{exp}$ (s)}}  & \multicolumn{1}{c|} {\textbf{Instr.}} & \multicolumn{1}{c|}
{\textbf{Grism}} & \multicolumn{1}{c|} {\textbf{airm.}} & \multicolumn{1}{c|}
{\textbf{Solar Analog (airm.)}} \\  \hline 
\endfirsthead
\multicolumn{8}{c}%
{{\bfseries \tablename\ \thetable{} -- continued from previous page}} \\ \hline 
\endfoot
\hline \multicolumn{1}{|c|} {\textbf{Object    }} & \multicolumn{1}{c|}
{\textbf{Night}} & \multicolumn{1}{c|} {\textbf{UT$_{start}$}} &
\multicolumn{1}{c|} {\textbf{T$_{exp}$ (s)}}  & \multicolumn{1}{c|} {\textbf{Instr.}} & \multicolumn{1}{c|}
{\textbf{Grism}} & \multicolumn{1}{c|} {\textbf{airm.}} & \multicolumn{1}{c|}
{\textbf{Solar Analog (airm.)}} \\  \hline 
\endhead
\hline \multicolumn{8}{r}{{Continued on next page}} \\ 
\endfoot
\hline \hline
\endlastfoot
24 Themis & 18 Dec. 2012 & 00:38 & 20 &  NICS & Amici & 1.03 & HIP22536 (1.02) \\
24 Themis & 18 Dec. 2012 & 02:16 & 10 &  Dolores & LRB & 1.22 & Lan98-978 (1.22) \\ 
24 Themis & 18 Dec. 2012 & 02:20 & 20 &  Dolores & LRR & 1.23 & Lan98-978 (1.22) \\ 
24 Themis & 18 Dec. 2012 & 21:33 & 10 &  Dolores & LRB & 1.14 & Lan115-271 (1.13) \\ 
24 Themis & 18 Dec. 2012 & 21:35 & 20 &  Dolores & LRR & 1.14 & Lan115-271 (1.13) \\ 
24 Themis & 19 Dec. 2012 & 01:07 & 60 &  Dolores & LRB & 1.07 & Lan115-271 (1.13) \\
62 Erato & 19 Feb. 2012 &  05:38    & 900  &  Dolores & LRR & 1.24 & Lan98-978(1.18) \\
90 Antiope & 18 Dec. 2012 & 01:48 & 30 &  Dolores & LRB & 1.20 & Lan102-1081 (1.15) \\ 
90 Antiope & 18 Dec. 2012 & 01:52 & 60 &  Dolores & LRR & 1.21 & Lan102-1081 (1.15) \\ 
90 Antiope & 17 Dec. 2012 & 23:29 & 60 &  NICS & Amici & 1.01 & HIP22536 (1.03) \\ 
268 Adorea & 20 Feb. 2012 &  06:08    & 300  &  Dolores & LRB & 1.33 & Hip59932 (1.32) \\
268 Adorea & 20 Feb. 2012 &  06:15    & 300  &  Dolores & LRR & 1.33 & Hip59932 (1.32) \\
383 Janina & 19 Feb. 2012 &  06:07    & 900  &  Dolores & LRR & 1.38 & Hip59932 (1.35) \\
461 Saskia & 19 Feb. 2012 &  03:52    & 900  &  Dolores & LRB & 1.27 & Hip59932 (1.34) \\ 
461 Saskia & 19 Feb. 2012 &  04:09    & 900  &  Dolores & LRR & 1.33 & Hip59932 (1.35) \\ 
461 Saskia & 19 Feb. 2012 &  23:44    & 720  &  NICS    & Amici & 1.09 & HIP44103 (1.04) \\ 
468 Lina   & 19 Dec. 2012 & 06:08 & 300 &  Dolores & LRR & 1.12 & Lan115-271 (1.14) \\ 
492 Gismonda & 18 Dec. 2012 & 22:45 & 300 &  Dolores & LRB & 1.56 & HIP52192 (1.46) \\ 
492 Gismonda & 18 Dec. 2012 & 22:51 & 300 &  Dolores & LRR & 1.59 & HIP52192 (1.47) \\
492 Gismonda & 16 Dec. 2012 & 20:47 & 480 &  NICS & Amici & 1.16 & Lan93-101 (1.14) \\
526 Jena & 19 Dec. 2012 & 00:23 & 300 &  Dolores & LRB & 1.17 &  HIP44027 (1.17) \\
526 Jena & 19 Dec. 2012 & 00:28 & 300 &  Dolores & LRR & 1.19 &  HIP44027 (1.17) \\ 
526 Jena & 17 Dec. 2012 & 22:54 & 240 &  NICS & Amici & 1.04 &  HIP22536 (1.03) \\ 
621 Werdandi  & 19 Feb. 2012 &  22:18    & 720 &  NICS    & Amici & 1.02 & HIP44103 (1.04) \\
656 Beagle & 18 Dec. 2012 & 20:02 & 600 &  Dolores & LRB & 1.23 & Lan93-101 (1.23) \\ 
656 Beagle & 18 Dec. 2012 & 20:14 & 600 &  Dolores & LRR & 1.26 & Lan115-271 (1.14) \\ 
656 Beagle & 17 Dec. 2012 & 19:39 & 960 &  NICS & Amici & 1.21 & Lan115-271 (1.25) \\ 
846 Lipperta & 20 Feb. 2012 &  03:19    & 900  &  Dolores & LRB & 1.21 & Lan102-1081(1.31) \\
846 Lipperta & 20 Feb. 2012 &  03:36    & 900  &  Dolores & LRR & 1.27 & Lan102-1081(1.31) \\
846 Lipperta & 20 Feb. 2012 &  00:19    & 960  &  NICS    & Amici & 1.06 & HIP44103 (1.04) \\ 
954  Li      & 19 Feb. 2012 &  21:23    & 1440 &  NICS    & Amici & 1.23 & Hyades64 (1.22) \\ 
1027 Aesculapia & 18 Dec. 2012 & 20:42 & 1200 &  Dolores & LRR & 1.33 & Lan115-271 (1.14) \\ 
1027 Aesculapia & 17 Dec. 2012 & 20:07 & 1920 &  NICS & Amici & 1.29 & Lan115-271 (1.25) \\ 
1247 Memoria & 19 Dec. 2012 & 05:04 & 900 &  Dolores & LRR & 1.04 & Lan115-271 (1.14) \\ 
1623 Vivian & 20 Feb. 2012 &  05:06    & 900  &  Dolores & LRB & 1.32 & Lan102-1081(1.31) \\
1623 Vivian & 20 Feb. 2012 &  05:24    & 900  &  Dolores & LRR & 1.39 & Lan102-1081(1.31) \\
1623 Vivian & 20 Feb. 2012 &  01:26    & 1440 &  NICS    & Amici & 1.12 & HIP44103(1.04) \\ 
1687 Glarona & 19 Feb. 2012 &  03:01    & 600  &  Dolores & LRB & 1.24 & Lan98-978 (1.18) \\ 
1687 Glarona & 19 Feb. 2012 &  03:17    & 600  &  Dolores & LRR & 1.28 & Lan98-978 (1.18) \\ 
1687 Glarona & 19 Feb. 2012 &  01:26    & 720 &  NICS    & Amici & 1.04 & HIP59932 (1.09) \\ 
1778  Alfven & 17 Dec. 2012 & 06:04 & 1440 & NICS & Amici & 1.29 & HIP41815 (1.26) \\ 
1953 Rupertwildt & 18 Dec. 2012 & 04:19 & 300 &  Dolores & LRB & 1.17 & Lan102-1081 (1.15) \\ 
1953 Rupertwildt & 18 Dec. 2012 & 04:25 & 480 &  Dolores & LRR & 1.19 & Lan102-1081 (1.15) \\ 
1953 Rupertwildt & 17 Dec. 2012 & 03:48 & 960 & NICS & Amici & 1.11 & HIP22536 (1.13) \\ 
2009 Voloshina & 17 Dec. 2012 & 23:58 & 960 & NICS & Amici & 1.24 & Lan98-978 (1.28) \\ 
2203 Van Rhijn & 17 Dec. 2012 & 21:24 & 960 & NICS & Amici & 1.05 & HIP22536 (1.03) \\ 
2222 Lermontov  & 20 Feb. 2012 &  04:36    & 900  &  Dolores & LRB & 1.34 & Lan102-1081(1.31) \\
2222 Lermontov  & 20 Feb. 2012 &  04:38    & 900  &  Dolores & LRR & 1.41 & Lan102-1081(1.31) \\
2222 Lermontov  & 20 Feb. 2012 &  00:48    & 960 &  NICS    & Amici & 1.05 & Lan102-1081 (1.18) \\ 
2228  Soyuz-Apollo& 19 Dec. 2012 & 04:30 & 600 &  Dolores & LRB & 1.03 &  Lan115-271 (1.13) \\ 
2228  Soyuz-Apollo & 19 Dec. 2012 & 04:42 & 600 &  Dolores & LRR & 1.03 &  Lan115-271 (1.14) \\
2264 Sabrina & 19 Dec. 2012 & 02:07 & 600 &  Dolores & LRB & 1.04 &  Lan115-271 (1.13) \\ 
2264 Sabrina & 18 Dec. 2012 & 06:21 & 600 &  Dolores & LRR & 1.45 &  Lan102-271 (1.15) \\ 
2264 Sabrina & 17 Dec. 2012 & 05:21 & 480 & NICS & Amici & 1.22 & Lan1021081 (1.16) \\ 
2270 Yazhi & 19 Dec. 2012 & 02:48 & 1100 &  Dolores & LRB & 1.01 &  Lan115-271 (1.13) \\ 
2270 Yazhi & 19 Dec. 2012 & 03:23 & 1100 &  Dolores & LRR & 1.02 &  Lan115-271 (1.13) \\ 
2270 Yazhi & 17 Dec. 2012 & 04:26 & 1920 &  NICS & Amici & 1.10 & HIP44027 (1.04) \\ 
2519    Annagerman & 19 Feb. 2012 &  04:46    & 900  &  Dolores & LRB & 1.16 & Land98-978 (1.17) \\ 
2519    Annagerman & 19 Feb. 2012 &  05:04    & 900  &  Dolores & LRR & 1.18 & Land98-978 (1.18) \\
2519    Annagerman & 20 Feb. 2012 &  02:13    & 1920 &  NICS    & Amici & 1.18 & HIP59932 (1.09) \\ 
3174 Alcock & 19 Dec. 2012 & 01:22 & 600 &  Dolores & LRB & 1.26 &  Lan93-101 (1.23) \\ 
3174 Alcock & 19 Dec. 2012 & 01:34 & 600 &  Dolores & LRR & 1.30 &  HIP41815 (1.30) \\ 
3174 Alcock & 16 Dec. 2012 & 23:52 & 960 & NICS & Amici & 1.06 & HIP22536 (1.13) \\ 
3591 Vladimirskij & 18 Dec. 2012 & 21:53 & 2200 &  Dolores & LRR & 1.21 &  HIP44027 (1.18) \\ 
3591 Vladimirskij & 16 Dec. 2012 & 22:20 & 1920 & NICS & Amici & 1.36 & Lan93-101 (1.14) \\ 
3615 Safronov & 16 Feb. 2012 & 21:31 & 900 & Dolores & LRB & 1.21 & Lan98-978 (1.25) \\ 
3615 Safronov & 16 Feb. 2012 & 21:49 & 900 & Dolores & LRR & 1.21 & Lan98-978 (1.25) \\ 
3615 Safronov & 18 Dec. 2012 & 05:47 & 1200 & Dolores & LRR & 1.13 & Lan102-1081 (1.15) \\ 
3615 Safronov & 17 Dec. 2012 & 02:40 & 480 & NICS & Amici & 1.05 & HIP41815 (1.06) \\ 
4903 Ichikawa & 18 Dec. 2012 & 02:53 & 900 & Dolores & LRB & 1.12 & Lan102-1081 (1.15) \\ 
4903 Ichikawa & 18 Dec. 2012 & 03:25 & 900 & Dolores & LRR & 1.15 & Lan102-1081 (1.15) \\ 
4903 Ichikawa & 17 Dec. 2012 & 01:38 & 960 & NICS & Amici & 1.02 & HIP22536 (1.13) \\ 
\hline
\hline
\end{longtable}
\end{center}
}

         \begin{sidewaystable} 
       \caption{Physical and orbital parameters of the Themis and Beagle families members investigated.  The albedo comes from WISE data (Masiero et al., 2011) when available, or from AKARI (indicated with the symbol $^*$, Usui et al., 2011). S$_{all}$ is the spectral slope value calculated 
in the whole {\bf observed} wavelength range, S$_{VIS}$ for the  0.55-0.80 $\mu$m range, S$_{NIR1}$ for the 0.9-1.4 $\mu$m range, S$_{NIR2}$
for the 1.4-2.2 $\mu$m range, and S$_{NIR3}$ for  1.1-1.8 $\mu$m range). The family is indicated with $B$ for the Beagle, and $T$ for the Themis members. For (24) Themis we used the whole visible and NIR spectrum acquired on 18 Dec., 00:30-02:30 UT time.}
        \label{slope}
        \scriptsize{
\begin{tabular}{|l|c|c|c|c|c|c|c|c|c|c|c|c|} \hline
Asteroid &   S$_{all}$         & S$_{Vis}$          & S$_{NIR1}$       & S$_{NIR2}$ & S$_{NIR3}$               &  D    & p$_{v}$ & Fam. & a & e & $sin(i)$ & H$_v$ \\          
         &  (\%/$10^{3}$\AA)   &  (\%/$10^{3}$\AA) & $(\%/10^{3}$\AA) & $(\%/10^{3}$\AA) & $(\%/10^{3}$\AA)  &  (Km) &         &      & (UA) &  &  & \\ \hline
 656        Beagle  &  1.30$\pm$0.50  &    -0.60$\pm$0.50  & 1.51$\pm$0.52 & 1.29$\pm$0.51 & 1.79$\pm$0.51& 47.58$\pm$0.33	& 0.0782$\pm$0.0222&   B &  3.15598& 0.154640 & 0.0235182&  9.95   \\ 
 1027    Aesculapia  &  0.56$\pm$0.50  &    -0.72$\pm$0.51  & 0.94$\pm$0.52 & 1.24$\pm$0.51 & 1.37$\pm$0.50& 35.38$\pm$0.34	& 0.0814$\pm$0.0071&   B &  3.15903& 0.152254 & 0.0240919& 10.82  \\ 
1687       Glarona  &  1.05$\pm$0.50  &    -1.36$\pm$0.51  & 1.95$\pm$0.53 & 1.69$\pm$0.52 & 2.32$\pm$0.51& 42.01$\pm$0.51  & 0.0795$\pm$0.0130&   B &  3.15870& 0.153173 & 0.0242712& 10.53  \\ 
2519    Annagerman  &  0.11$\pm$0.50  &    -0.65$\pm$0.51  &-0.56$\pm$0.51 & 0.60$\pm$0.54 & 0.62$\pm$0.55& 20.55$\pm$0.31  & 0.1152$\pm$0.0289&   B &  3.14765& 0.151787 & 0.0225837& 11.5    \\ 
3174        Alcock  &  0.19$\pm$0.50  &    -2.88$\pm$0.50  & 1.37$\pm$0.51 & 0.39$\pm$0.52 & 1.73$\pm$0.50& 18.66$\pm$0.80	& 0.1020$\pm$0.0090&   B &  3.15479& 0.154074 & 0.0239525& 11.95  \\
3591  Vladimirskij  & -0.57$\pm$0.50  &    -3.24$\pm$0.51  & 0.16$\pm$0.52 & 0.88$\pm$0.52 & 0.91$\pm$0.51& 16.42$\pm$0.25	& 0.0947$\pm$0.0248&   B &  3.15704& 0.151941 & 0.0244373& 12.07   \\ 
3615      Safronov  & -0.79$\pm$0.51  &    -1.63$\pm$0.50  &-3.06$\pm$1.30 & 1.85$\pm$1.16 &-0.17$\pm$0.94& 26.24$\pm$0.10	& 0.0852$\pm$0.0159&   B &  3.15974& 0.152726 & 0.0238029& 11.49   \\ 
4903      Ichikawa  &  0.40$\pm$0.50  &    -1.56$\pm$0.51  & 1.51$\pm$0.55 & 0.77$\pm$0.62 & 2.11$\pm$0.52& 14.79$\pm$0.39	& 0.1167$\pm$0.0087&   B &  3.15055& 0.15180  & 0.0240145& 12.54  \\ \hline
  24       Themis   &  0.97$\pm$0.51  &    -0.61$\pm$0.50  & 1.24$\pm$0.51 & 1.95$\pm$0.50 & 2.15$\pm$0.50& 202.34$\pm$6.05 & 0.0641$\pm$0.0157&   T &  3.13450& 0.152779 & 0.0189278&  7.21  \\ 
  62         Erato  & -0.06$\pm$0.51  &    -1.73$\pm$0.50  &-0.23$\pm$0.50 & 0.73$\pm$0.50 & 0.32$\pm$0.50& 95.40$\pm$2.0   & 0.0610$\pm$0.0030&   T &  3.12169& 0.149807 & 0.0225411&  8.62   \\ 
  90       antiope  &  0.88$\pm$0.50  &    -1.19$\pm$0.50  & 1.45$\pm$0.51 & 2.08$\pm$0.51 & 2.38$\pm$0.51& 121.13$\pm$2.47	& 0.0593$\pm$0.0096&   T &  3.14619& 0.153819 & 0.0231415&  7.84  \\  
268         Adorea  &  2.55$\pm$0.50  &     1.56$\pm$0.50  & 1.29$\pm$0.51 & 3.76$\pm$0.50 & 3.42$\pm$0.50&140.59$\pm$3.18  & 0.0436$\pm$0.0066&   T &  3.09684& 0.167397 & 0.0243772&  8.3      \\
  383        Janina  &  0.61$\pm$0.50  &     2.97$\pm$0.51  & 0.59$\pm$0.51 & 0.42$\pm$0.50 & 0.48$\pm$0.50& 44.64$\pm$0.74  & 0.0964$\pm$0.0127&   T &  3.13408& 0.151863 & 0.0246743&  9.74   \\ 
 461        Saskia  &  3.19$\pm$0.50  &     2.79$\pm$0.50  & 2.49$\pm$0.51 & 3.04$\pm$0.51 & 3.50$\pm$0.51& 48.72$\pm$0.24  & 0.0479$\pm$0.0079&   T &  3.11426& 0.155859 & 0.0239453& 10.4     \\ 
468           Lina  &   -             &    -1.41$\pm$0.53  &  -            &  -            &  -           & 64.59$\pm$1.98  & 0.0495$\pm$0.0094&   T &  3.14189& 0.158666 & 0.0215053&  9.69 \\
 492      Gismonda  &  0.79$\pm$0.50  &    -1.42$\pm$0.50  & 0.84$\pm$0.51 & 1.21$\pm$0.51 & 1.22$\pm$0.51& 59.92$\pm$0.35  & 0.0592$\pm$0.0047&   T &  3.11191& 0.151069 & 0.0220835&  9.85   \\ 
 526          Jena  & -0.26$\pm$0.50  &    -0.95$\pm$0.50  &-1.60$\pm$0.51 &-0.63$\pm$0.50 &-0.25$\pm$0.51& 51.03$\pm$0.74  & 0.0580$\pm$0.0177&   T &  3.12313& 0.156463 & 0.0249680& 10.      \\ 
 621      Werdandi  &  1.05$\pm$0.50  &    -0.51$\pm$0.51  & 2.01$\pm$0.51 & 1.98$\pm$0.52 & 1.91$\pm$0.51& 30.71$\pm$0.50$^{*}$   & 0.1240$\pm$0.0050$^{*}$ &   T &  3.1193 & 0.150097 & 0.0253958& 10.98    \\ 
 846      Lipperta  &  1.39$\pm$0.50  &     0.50$\pm$0.50  & 1.72$\pm$0.51 & 2.11$\pm$0.51 & 2.83$\pm$0.51& 51.45$\pm$0.76$^{*}$	& 0.0530$\pm$0.0020$^{*}$&   T &  3.12791& 0.150758 & 0.0268551& 10.25    \\ 
 954            Li  &  2.83$\pm$0.50  &     1.26$\pm$0.50  & 3.00$\pm$0.53 & 3.06$\pm$0.53 & 4.15$\pm$0.53& 52.06$\pm$0.81  & 0.0552$\pm$0.0073&   T &  3.13701& 0.150656 & 0.0230923& 10.07    \\ 
1247       Memoria  &   -             &    -0.64$\pm$0.51  &  -            &  -            &  -           & 42.07$\pm$0.16  & 0.0618$\pm$0.0060&   T &  3.13875& 0.160209 & 0.0297064& 10.58 \\
 1623        Vivian  &  2.53$\pm$0.50  &     1.62$\pm$0.50  & 1.87$\pm$0.51 & 2.62$\pm$0.52 & 3.20$\pm$0.52& 29.98$\pm$1.74$^{*}$	& 0.0780$\pm$0.0100$^{*}$ &   T &  3.13466& 0.15294  & 0.0239821& 11.19     \\
 1778        Alfven &   0.17$\pm$0.50  &       --           & 0.14$\pm$0.51 & -0.47$\pm$0.53 & 0.29$\pm$0.50 & 20.6$\pm$0.24 & 0.0951$\pm$0.0069 &  T   & 3.1441 &  0.155394 &  0.0224451 & 11.7 \\ 
 1953   Rupertwildt  &  0.55$\pm$0.50  &    -1.60$\pm$0.50  & 0.79$\pm$0.51 & 0.89$\pm$0.52 & 2.16$\pm$0.51& 21.97$\pm$0.27  & 0.0697$\pm$0.0060&   T &  3.11586& 0.147538 & 0.0242019& 11.84   \\ 
2009     Voloshina  &  1.78$\pm$0.50  &      -             & 1.43$\pm$0.51 & 1.71$\pm$0.51 & 2.35$\pm$0.50& 26.56$\pm$0.48  & 0.1199$\pm$0.0242 &   T &  3.11914 & 0.150115 & 0.0282294 & 11.16 \\
2203     vanRhijn   & 1.39$\pm$0.50   &      --            &  1.41$\pm$0.51&  1.00$\pm$0.51 & 1.90$\pm$0.51 & 22.28$\pm$0.27 & 0.0898$\pm$0.0058 &  T &  3.10989 & 0.149153 &  0.0209356 & 11.57 \\
2222     Lermontov  &  2.39$\pm$0.50  &     1.96$\pm$0.50  & 2.05$\pm$0.51 & 2.87$\pm$0.52 & 2.85$\pm$0.51& 32.22$\pm$0.22  & 0.0469$\pm$0.0061&   T &  3.11619& 0.159625 & 0.0228687& 11.29     \\
2228  Soyuz-Apollo  &   -             &     1.65$\pm$0.51  &  -            &  -            &  -           & 26.08$\pm$0.29  & 0.1134$\pm$0.0198&   T &  3.14342& 0.160273 & 0.0241862& 11.44  \\
 2264       Sabrina  &  0.08$\pm$0.50  &    -1.49$\pm$0.50  & 0.44$\pm$0.52 & 0.40$\pm$0.51 & 1.56$\pm$0.51& 37.34$\pm$0.38  & 0.0799$\pm$0.0138&   T &  3.13084& 0.154135 & 0.0234201& 10.92   \\ 
2270         Yazhi  &  1.82$\pm$0.50 &    -0.07$\pm$0.51  & 1.12$\pm$0.55 & 0.06$\pm$0.54 & 1.65$\pm$0.52& 26.73$\pm$0.30  & 0.1081$\pm$0.0225&   T &  3.15117& 0.15233  & 0.0188940& 11.43    \\ 
 \hline
\end{tabular}
}
         \end{sidewaystable}

         \thispagestyle{empty}

 \begin{table}
         \begin{center} 
         \caption{Identification of absorption bands attributed to hydrated silicates on the Themis family members.}
        \label{hydra}
                \begin{tabular}{|l|c|c|c|c|} \hline
         Asteroid     & W$_{in}$ (\AA) & W$_{fin}$ (\AA) & Band$_{center}$ (\AA) & Depth (\%) \\ \hline
         24 Themis       & 5715  &  8730  &   7333$\pm$48 &   2.6$\pm$0.1  \\
         90 Antiope      & 5358  &  8375  &   6972$\pm$43  & 1.4$\pm$0.1  \\
         461 Saskia      & 5648  &  8338  &   6881$\pm$123  & 1.2$\pm$0.1  \\
     	                 & 12397  &  15126  &  13175$\pm$55  & 2.8$\pm$0.2  \\
                         & 15965  &  21378  &  18157$\pm$31  & 2.7$\pm$0.2  \\
         846 Lipperta     & 5487  &  8480  &    6842$\pm$65  & 1.02$\pm$0.1 \\ \hline
\end{tabular}
\end{center}
\end{table}

 \newpage
 
 \begin{table}
         \begin{center} 
         \caption{RELAB matches. The first eight asteroids belong to the Beagle family.}
         \label{met}
          \scriptsize{
        \begin{tabular}{|l|c|c|c|c|c|} \hline
          Asteroid & Albedo & Best fit & Met refl. & Met Class \& name &  Grain size \\ \hline
 656           Beagle  & 0.0782   &     NAMC02      & 0.033 &  CM2 LEW90500,45 &  $<$ 500 $\mu$m           \\
 1027        Aesculapia  & 0.0814 &     C1MP61      & 0.067 &  CM2 GRO95577,6 particulate & $<$ 125 $\mu$m           \\
 1687           Glarona  & 0.0795 &     camh52      & 0.068 &  CM2  Boriskino  & $<$ 45 $\mu$m         \\
2519       Annagerman  & 0.1152   &     c1mb20      & 0.063 &  CM2  Unusual Y-86720,77 &    $<$ 125 $\mu$m            \\
 3174            Alcock  & 0.102  &     MGP094      & 0.059 &  CM2 Murchison          & whole rock  \\ 
3591      Vladimirskij  & 0.0947  &     c1mb19      & 0.036 &  CI  unusual, Y-82162,79 & $<$ 150 $\mu$m     \\
3615          Safronov  & 0.0852  &     BKR1MA076   & 0.042 &  CM2/CR Kaidun            & $> $  250 $\mu$m        \\
4903          Ichikawa  & 0.1167 &      CGP090      & 0.046 &  CM2 Cold Bokkevelt      & 150 -- 500 $\mu$m    \\   \hline
  24   Themis (17 Dec)  & 0.0641 &      NAMC02      & 0.033 &  CM2 LEW90500,45         & $<$ 500 $\mu$m           \\
  62             Erato  & 0.061  &      ccmb19      & 0.050 &  CI unusual, Y-82162,79  &    \\
  90           antiope  & 0.0593 &      NAMC02      & 0.033 &  CM2 LEW90500,45         & $<$ 500 $\mu$m           \\
  268           Adorea  & 0.0436 &      ccmb64      & 0.031 &  CM2 Murchison heated 600 C & $<$ 63 $\mu$m \\
 383            Janina  & 0.0964 &      c1mb18      & 0.06  &  CM2 unusual B-7904,108     &             \\ 
 461            Saskia  & 0.0479 &      s1rs42      & 0.037 &  CM2 Mighei                 &   \\
 492          Gismonda  & 0.0592 &      C1RS46      & 0.036 &  CM2 Boriskino             & \\
 526              Jena  & 0.0580 &      CGP130      & 0.047 &  CV3 Grosnaja             & 75--150$\mu$m             \\
 621          Werdandi  & 0.124  & BKR1MP022L0      & 0.046 &  CM2 MAC88100  laser irr.             & $<$ 125 $\mu$m       \\    
 846          Lipperta  &  0.053 &      ncmp22      & 0.033 &  CM2 MAC88100,30           & $<$ 125 $\mu$m \\
 954                Li  & 0.0552 &      s1rs42      & 0.035 &  CM2 Mighei               &  \\
1623            Vivian  &  0.078 &      ccmb64      & 0.031 &  CM2 Murchison heated 600 C &  $<$ 500 $\mu$m     \\
 1953       Rupertwildt  & 0.0697 &     CGP142      & 0.046 &  CM2 ColdBokkevelt         & 75-500 $\mu$m\\
2222         Lermontov  & 0.0469 &      ccmb64      & 0.031 &  CM2 Murchison heated 600 C	& $<$ 500 $\mu$m  \\
2264           Sabrina  & 0.0799 &      CGP094      & 0.057 &  CM2 Murchison            & whole rock\\
2270             Yazhi  & 0.1081 &      CBMS01      & 0.035 &  CM2 Migei Separates      & 100 -- 200 $\mu$m   \\ \hline
        \end{tabular}
        }
\end{center}
 \end{table}

\end{document}